\documentclass[]{aa}
\usepackage{graphics,latexsym,epsfig,graphicx}     

\newcommand{\dt}[1]{\frac{\partial #1}{\partial t}}
\newcommand{\pow}[1]{P_{\rm #1}}
\newcommand{\ave}[1]{\big\langle #1 \big\rangle}
\newcommand{\avp}[1]{\big\langle #1 \big\rangle^\prime}
\newcommand{\Ave}[1]{\Big\langle #1 \Big\rangle^\prime}
\newcommand{\Ref}[1]{(\ref{#1})}

\newcommand{\Vector}[1]{{\bf #1}}
\newcommand{\Matrix}[1]{{\bf #1}}
\newcommand{\pprime}[0]{{\prime\prime}}
\renewcommand{\Re}[0]{{\rm Re~}}
\begin{document}

\title{Time evolution of the stochastic linear bias of interacting
  galaxies on linear scales} \author{Patrick Simon}

\institute{ Institut f\"{u}r Astrophysik und Extraterrestrische
  Forschung, Auf dem H{\"u}gel 71, D-53121 Bonn, Germany }

\date{} \authorrunning{Patrick Simon} \titlerunning{Bias model for
  interacting galaxies}

\abstract{ We extend the time dependent bias model of
  Tegmark~\&~Peebles (1998) to predict the large-scale evolution of
  the stochastic linear bias of different galaxy populations with
  respect to both the dark matter and each other. The resulting model
  equations contain a general expression, coined the ``interaction
  term'', accounting for the destruction or production of galaxies.
  This term may be used to model couplings between different
  populations that lead to an increase or decrease of the number of a
  galaxies belonging to a population, e.g. passive evolution or
  merging processes. This is explored in detail using a toy model.  In
  particular, it is shown that the presence of such a coupling may
  change the evolution of the bias parameter compared to an
  interaction-free evolution.  We argue that the observation of the
  evolution of the large-scale bias and galaxy number density with
  wide-field surveys may be used to infer fundamental interaction
  parameters between galaxy populations, possibly giving an insight in
  their formation and
  evolution.}
\maketitle
\section{Introduction}

The first studies of large scale structure had to completely rely on
galaxies as mass tracers of the large scale Universe. It became clear
that in order to match the clustering statistics of galaxies with
models for gravitationally driven structure formation - in particular
with the SCDM favoured in the early 90s - galaxies cannot be perfect
tracers of the overall mass density field; the concept of galaxy
biasing was born (e.g. Bardeen et al. 1986, BBKS hereafter; Davis et
al. 1985).

The first bias description was introduced by Kaiser (1984) as a single
parameter that rescales the two-point correlation function (2PCF) of
the galaxy density field to yield the expected 2PCF for matter
clustering. Such a rescaling can be achieved if the fluctuation field
or density contrast $\delta_{\rm m}\equiv\rho_{\rm m}/\bar{\rho}_{\rm
  m}-1$ of the matter density field is a linear function of the galaxy
density contrast $\delta_{\rm g}\equiv{\rm n}_{\rm g}/\bar{\rm n}_{\rm
  g}-1$, thus $\delta_{\rm b}=b\delta_{\rm m}$. $\rho_{\rm m}$ and
${\rm n}_{\rm g}$ are the matter and galaxy number density fields
respectively. The bar denotes the mean density. A possible reason for
the enhancement of clustering could be that galaxies are
preferentially formed in the peaks of the dark matter field (Kaiser
1984; BBKS).

This linear biasing scheme was put in a more general framework by
Fry~\&~Gazta\'{n}aga (1993) who proposed $\delta_{\rm g}$ to be an
arbitrary analytic local function of $\delta_{\rm m}$ (local Eulerian
biasing), opening the door to (in principal arbitrarily many) bias
parameters which, however, can be measured if higher-order statistics
is invoked.  Moreover, these parameters can be different if different
smoothing scales of the density fields are considered
(scale-dependence of bias).  An alternative picture to the Eulerian
model is to assume that the galaxy distribution or a part of it (like
recently formed galaxies) is only a local function of the dark matter
field at one particular time (Lagrangian bias: e.g.  Catelan et al.
2000).

In order to look at general features of the statistics of transformed
random fields, Coles (1993) derived constraints for the clustering of
galaxies that follow from a local mapping of a Gaussian field.  It was
found that on large scales where clustering is small - even for a
non-local mapping that, however, preserves the clustering hierarchy
(Coles, Melott~\&~Munshi 1999; Scherrer~\&~Weinberg 1998) - the shape
of the 2PCF of the matter and the galaxies is identical, so that here
a simple linear bias scheme may still be used (see also Narayanan,
Berlind~\&~Weinberg 2000 and Mann, Peacock~\&~Heavens 1998).

Another degree of freedom had to be inserted into the biasing schemes,
once it was realised that the relation between the matter and the
galaxy density field is very likely to be a stochastic one (Blanton
2000; Tegmark~\&~Bromley 1999; Dekel~\&~Lahav 1999; Matsubara 1999;
Cen~\&~Ostriker 1992) due to ``hidden parameters'' of galaxy
formation/evolution that cannot be incorporated into a simple picture
that involves only the densities. Currently, the description for
stochastic nonlinear biasing most commonly used is by Dekel~\&~Lahav
(1999). It expresses the joint probability distribution
$P\left(\delta_{\rm g},\delta_{\rm m}\right)$ of local values for
$\delta_{\rm g}$ and $\delta_{\rm m}$ in terms of the conditional mean
$\ave{\delta_{\rm g}|\delta_{\rm m}}$ and the scatter $\delta_{\rm
  g}-\ave{\delta_{\rm g}|\delta_{\rm m}}$ about the conditional mean.
For $P\left(\delta_{\rm g},\delta_{\rm m}\right)$ being a bivariate
Gaussian this scheme reduces to the following stochastic linear
biasing parameters
\begin{equation}\label{linearbias2}
  b\equiv\sqrt{\frac{\ave{\delta^2_{\rm g}}}{\ave{\delta^2_{\rm m}}}}
  \;\;;\;\;
  r\equiv\frac{\ave{\delta_{\rm m}\delta_{\rm g}}}
    {\sqrt{\ave{\delta^2_{\rm g}}\ave{\delta^2_{\rm m}}}}
  \; ,
\end{equation}
which is approximately fulfilled on large scales. Equivalently, one
can express the bias parameters in terms of auto- and
cross-correlation power spectra $P_{\rm m}=\ave{\tilde{\delta}_{\rm
    m}\tilde{\delta}_{\rm m}^\ast}^\prime$, $P_{\rm
  gg}=\ave{\tilde{\delta}_{\rm g}\tilde{\delta}_{\rm g}^\ast}^\prime$
and $P_{gm}=\ave{\tilde{\delta}_{\rm m}\tilde{\delta}_{\rm
    g}^\ast}^\prime$:
\begin{equation}\label{linearbias}
  b\left(k\right)\equiv\sqrt{\frac{P_{\rm gg}\left(k\right)}
    {P_{\rm m}\left(k\right)}}
  \;\;\;\;
  r\left(k\right)\equiv\frac{P_{\rm gm}\left(k\right)}
    {\sqrt{P_{\rm gg}\left(k\right)P_{\rm m}\left(k\right)}}
  \; ,
\end{equation}
where $\tilde{\delta}_{\rm g}$ and $\tilde{\delta}_{\rm m}$ are the
Fourier transforms of the density contrasts. The asterisk ``$\ast$''
denotes the complex conjugate\footnote{For the definition of
  $\ave{...}^\prime$ kindly see Eq. \Ref{defaveprime} in Sect. 2.3.}.
For the Fourier space definition above, the relevant scale for the
bias parameters is defined by the spatial frequency considered,
whereas in the other definition a scale is defined by smoothing the
fields with a kernel of a certain size.  In the case of a strict
positive correlation $r=1$, the prescription reduces to the former
deterministic linear bias with only one free parameter $b$.

Nowadays, in a time of a ``concordance'' cosmological model (Spergel
et al. 2003), cosmic microwave background and weak gravitational
lensing studies provide information on the matter distribution almost
independent of the galaxy distribution, confirming the original
paradigm of structure formation, in particular the $\Lambda$CDM model.
Conversely, this also confirms the early suspicion that galaxies are
not perfect mass tracers. Intriguingly, in contrast to about ten years
ago when SCDM was the favoured cosmological model, with $\Lambda$CDM
almost no bias in the local Universe on large scales is required
(Verde et al. 2002).  However, there is a need for bias on smaller
scales (scale-dependent bias), because in contrast to that of the dark
matter the 2PCF of the galaxies is a power law over several orders of
magnitude. The exact scale-dependence of the bias maybe even hold some
information on the physics of galaxy formation (e.g.  Benson et al.
2000; Blanton et al.  1999).

The evolution of bias has also become a matter of interest: there is
evidence that galaxy clustering is a function of redshift (e.g.
Carlberg et al. 2000; Adelberger et al. 1998; Le Fevre et al. 1996)
and even that the galaxy bias is a decreasing function of redshift
(e.g.  Blanton et al. 2000; Magliocchetti et al. 2000; Steidel et al.
1998; Wechsler et al. 1998; Matarrese et al. 1997).

Analytical models for the bias evolution fall into two categories:
test particle models and halo models. Test particle models
(Basilakos~\&~Plionis 2001; Matsubara 1999; Taruya~\&~Soda 1999;
Taruya et al. 1999; Tegmark~\&~Peebles 1998, hereafter TP98; Fry 1996,
hereafter F96; Nusser~\&~Davis 1994) assume that galaxies passively
follow the bulk flow of the dark matter field which then can be
treated with conventional perturbation theory.  Halo models (Seljak
2000; Peacock~\&~Smith 2000; Sheth~\&~Lemson 1999; Bagla 1998; Catelan
et al. 1998; Mo~\&~White 1996), on the other hand, picture the dark
matter density field to be made up out of typical haloes that host
galaxies, so that the clustering of galaxies is related to the
clustering of their hosts and typical halo properties. Both concepts
agree on a debiasing of the galaxy field with time, but there are
differences in the details (Magliocchetti et al. 2000).

If one does not look at the galaxy population as a whole, which as
noted above seems to trace the local matter distribution quite well on
large scales, there are more interesting features.  It is also known
that different types of galaxies are differently clustered with
respect to each other, and, consequently, also with respect to the
underlying dark matter field (e.g. Phleps~\&~Meisenheimer 2003 and
references therein; Norberg et al. 2002; Blanton et al.  2000). At low
redshift, the correlation length - a measure for the strength of
clustering - is a function of morphological type and color (Tucker et
al. 1997; Loveday et al. 1995; Davis~\&~Geller 1976) and maybe also
depend on the luminosity of the galaxy population (Benoist et al.
1996).  Furthermore, there are examples of galaxy populations whose
relative clustering is known to have changed with time. For instance,
red and blue galaxies were almost not biased with respect to each
other at $z\ge 0.5$ (Le Fevre et al.  1996; but also see
Phleps~\&~Meisenheimer 2003 who do not observe this bias), but today
early type galaxies are more strongly clustered than late types (e.g.
Norberg 2002; Baker et al. 1998). 

 Thus, it makes sense to conceive a model for bias
evolution that takes into account several distinct galaxy populations.

Observationally, the stochastic linear bias can be measured by
redshift space distortions (Sigad, Branchini~\&~Dekel 2000; Pen 1998;
Kaiser 1987), weak gravitational lensing (Fan 2003; Hoekstra et al.
2002; van Waerbeke 1998; Schneider 1998) and counts-in-cells
statistics (Conway et al. 2004; Tegmark~\&~Bromley 1999; Efstathiou et
al. 1990); the latter, however, only for biasing between galaxies.
Future surveys with an appropriate number of galaxies will be required
to obtain a good signal-to-noise ratio for reconstruction of the
evolution of bias.

In this paper, we extend the test particle model of TP98 for the
stochastic linear parameter evolution and include several galaxy
populations that are allowed to interact with each other. The rate of
galaxy interaction is assumed to be a function of all density fields,
changing in general the number of members of a particular galaxy
population. Treated is also the evolution of the relative bias of the
populations with respect to each other, not only the bias relative to
the dark matter field.

In detail, the second section develops a model based on the bulk flow
hypothesis including a general sink/source term for galaxies.  We
derive differential equations for the auto- and cross-correlation
power spectra (galaxy-galaxy, galaxy-dark matter), valid on scales
where the fields are Gaussian, thus on linear scales (Sect. 2.3).  The
equations are then transformed to obtain differential equations for
the stochastic linear bias parameters (Sect. 2.4). In Sect. 3, we
focus on linear and quadratic interaction rates and work out the
relevant terms needed for the bias model equations based on this
interaction (table \ref{tab}).  We demonstrate in Sect. 4 for a few
toy models the effect on the evolution of the large scale bias in the
presence of galaxy interactions.  We conclude this paper with a
discussion.


\section{Derivation of the bias model}

\subsection{Evolution of density contrasts}

Here we derive differential equations for the density contrasts of a
set of galaxy species that are assumed to be perfect velocity tracers,
meaning that their bulk velocities are identical to the overall bulk
mass flow.

It is common practice to express the density fields - dark matter
$\rho_{\rm m}$ and galaxies $n_i$, $i=1..N$ - in terms of the their
mean density and the fluctuation about this mean, the \emph{density
  contrast}:
\begin{eqnarray}\label{densitycontrast}
  \rho_{\rm m}=\bar{\rho}_{\rm m}\left(1+\delta_{\rm m}\right)
  &&
  n_i=\bar{n}_i\left(1+\delta_i\right)
  \; .
\end{eqnarray}
$\bar{\rho}_{\rm m}\equiv\langle\rho_{\rm m}\rangle$ and
$\bar{n}_i\equiv\langle n_i\rangle$ are the corresponding matter and
number density respectively, obtained by taking the volume average.
Under the usual hydrodynamic conditions and cosmological assumptions
(Peebles 1980), the gravitationally driven evolution of the dark
matter density and bulk flow $\Vector{v}$ (deviation from the Hubble
flow) is described in comoving coordinates by
\begin{eqnarray}
  \nonumber
  \frac{\partial\theta}{\partial
  t}+\left(1-\frac{\Omega_{\rm m}}{2}+\Omega_\Lambda\right)H\left(a\right)\theta&&
  \\\label{hydro2}
  +\frac{3}{2}H\left(a\right)\Omega_{\rm m}\delta_{\rm m}+\frac{1}{a^2H\left(a\right)}
  \nabla\left(\Vector{v}\nabla\right)\Vector{v}&=&0
  \\\label{hydro1}
  \frac{\partial\delta_{\rm m}}{\partial t}+H\left(a\right)\theta+
  \frac{1}{a}\nabla\left(\delta_{\rm m}\Vector{v}\right) &=& 0
\end{eqnarray}
\begin{equation}\label{cosmology}
  H\left(a\right)=
  H_0\sqrt{\Omega_{\rm m}
  a^{-3}+\left(1-\Omega_{\rm m}-\Omega_\Lambda\right)a^{-2}+\Omega_\Lambda}
  \; .
\end{equation}
$\theta\equiv \nabla\Vector{v}/\left[aH\left(a\right)\right]$ is
related to the curl free component of the bulk flow. On large scales,
this is the only component (in contrast to the vorticity) that is not
suppressed by structure growth and therefore the only one relevant for
structure formation. $H\left(a\right)\equiv \dot{a}/a$ is the Hubble
parameter. Solutions to Eqs.  \Ref{hydro2}-\Ref{cosmology} have been
extensively studied in the literature, especially using the
perturbation approach (e.g.  Bernardeau et al. 2001 for a review;
Goroff et al. 1986) and therefore will not be discussed here.  We
simply assume that the solutions for $\delta_{\rm m}$ and $\Vector{v}$
or $\theta$ are (approximately) known.

The central assumption in this and similar models (e.g. TP98; F96) is
that the velocity fields of the galaxies are identical to that of the
dark matter. One thereby reduces the treatment for the galaxy number
density solely to the number conservation equation, which for a
\emph{conserved number of galaxies} looks as Eq. \Ref{hydro1} (TP98):
 \begin{equation}\label{hydro3}
  \frac{\partial\delta_i}{\partial t}+H\left(a\right)\theta+
  \frac{1}{a}\nabla\left(\delta_i\Vector{v}\right)=0
  \; .
\end{equation}
The term $H\left(a\right)\theta$ can be removed by subtracting Eq. \Ref{hydro1},
arriving at an equation that clearly shows how the galaxies are
coupled to the dark matter field
\begin{equation}\label{hydro4}
  \frac{\partial\delta_i}{\partial t}=
  \frac{\partial\delta_{\rm m}}{\partial t}+
  \frac{1}{a}\nabla\left(\Vector{v}\left[\delta_{\rm m}-\delta_i\right]\right)
  \; .
\end{equation}
Our main modification consists of dropping the constraint that the
mean number of galaxies - expressed by $\bar{n}_i$ - is conserved.  We
allow for a sink/source term $\Phi_i$ in the mass conservation
equation for the galaxy population $n_i$ that incorporates
galaxy-galaxy and galaxy-dark matter interactions, and is thought to
be a function of all the density fields. Note that in this formalism
interaction is equivalent to a change in galaxy number density.

In order to include $\Phi_i$ in Eq. \Ref{hydro3} and to eventually
obtain a modified formula \Ref{hydro4}, we have to start with the
number conservation equation for the galaxies plus the new interaction
rate $\Phi_i$:
\begin{equation}
 \frac{\partial n_i}{\partial
 t}+\frac{1}{a}\nabla\left(\Vector{v}n_i\right)=\Phi_i
 \; .
\end{equation}
Setting $\Phi_i=0$ would result again in Eq. \Ref{hydro4}.
Substitution of $n_i$ by the definition in \Ref{densitycontrast}
yields:

\begin{equation}
\frac{\partial\delta_i}{\partial t}+
H\left(a\right)\theta+
\frac{1}{a}\nabla\left(\Vector{v}\delta_i\right)
=
\frac{1}{\bar{n}_i}
\left[\Phi_i-\left(1+\delta_i\right)\frac{\partial\bar{n}_i}{\partial
  t}\right]
\; .
\end{equation}
For the last step we had to take into account that the mean galaxy
density $\bar{n}_i$ is a function of time.  Compared to Eq.
\Ref{hydro3}, we obviously have a new term on the right hand side
(rhs) that has to be cared for. Again, subtracting Eq. \Ref{hydro1}
from the last equation gives the time evolution equation for the
density contrasts of the galaxies but this time accounting for the
impact of a varying mean galaxy density due to $\Phi_i$
\begin{eqnarray}\label{basiceq}
&&\frac{\partial\delta_i}{\partial t}=
\\\nonumber
&&\frac{\partial\delta_{\rm m}}{\partial t}+
\frac{1}{a}\nabla\left(\Vector{v}\left[\delta_{\rm m}-\delta_i\right]\right)
+\frac{1}{\bar{n}_i}
\left[\Phi_i-\delta_i
\frac{\partial\bar{n}_i}{\partial t}\right]
-\frac{1}{\bar{n}_i}\frac{\partial\bar{n}_i}{\partial t}
\; .
\end{eqnarray}

\subsection{Evolution of mean densities}

In order to get the time-dependence of the mean galaxy density 
$\bar{n}_i$, we take the ensemble average\footnote{Due to the ergodicity
  of the random fields involved, volume and ensemble average are
  identical.}  $\ave{...}$ of Eq. \Ref{basiceq}:
\begin{equation}\label{lindensity}
  \frac{\partial\bar{n}_i}{\partial
    t}=\ave{\Phi_i}
 \; ,
\end{equation}
where we used
$\ave{\delta_i}=\ave{\delta_m}=\ave{\nabla\left(\Vector{v}\delta_i\right)}=\nabla\ave{\Vector{v}\delta_i}=0$.

The terms $\nabla\ave{\Vector{v}\delta_{\rm m}}$ and
$\nabla\ave{\Vector{v}\delta_i}$ vanish, because the net flux
\begin{equation}
  \ave{\rho\Vector{v}}=
  \bar{\rho}\ave{\Vector{v}\delta}+\bar{\rho}\ave{\Vector{v}}=0
\end{equation}
of any species $\rho$ over the whole volume has to be zero, since we
work in the rest frame of the Hubble expansion.  In particular, Eq.
\Ref{lindensity} has general validity and is not restricted to
Gaussian fields only.

\subsection{Linear scale evolution of correlation power spectra}

We will primarily be interested in the evolution of the linear
stochastic bias which may be expressed in terms of the cross- and
auto-correlation power spectra. Therefore, the next logical step is to
work out the time dependence of these power spectra. For that reason,
we take the Fourier transform
\begin{equation}\label{fouriertrans}
  \tilde{\delta}\left(\Vector{k}\right)=
  \frac{1}{\left(2\pi\right)^3}\int
  d^3\Vector{r}~\delta\left(\Vector{r}\right){\rm
  e}^{-{\rm i}\Vector{k}\Vector{r}}
\end{equation}
of Eq. \Ref{basiceq} to obtain the corresponding equation for the
Fourier coefficients
\begin{eqnarray}\label{basiceqfourier}
&&\frac{\partial\tilde{\delta}_i}{\partial t}=
\\\nonumber
&&
\frac{\partial\tilde{\delta}_{\rm m}}{\partial t}+
\frac{1}{\bar{n}_i}
\left[\tilde{\Phi}_i-\tilde{\delta}_i
\frac{\partial\bar{n}_i}{\partial t}\right]
+\frac{\rm i~\Vector{k}}{a}\left(\Vector{\tilde{v}}
\ast\left[\tilde{\delta}_{\rm m}-\tilde{\delta}_i\right]\right)
\; , 
\end{eqnarray}
where the irrelevant terms at $\Vector{k}=0$ have been neglected.  For
convenience, we omit the arguments in the brackets of the Fourier
coefficients.  By the asterisk ``$\ast$'' we denote the convolution of
two fields in Fourier space
\begin{equation}
  \left(\tilde{f}\ast\tilde{g}\right)\left(\Vector{k}\right)\equiv
  \frac{1}{\left(2\pi\right)^3}\int
  d^3\Vector{k}^\prime\tilde{f}\left(\Vector{k}^\prime\right)
  \tilde{g}\left(\Vector{k}-\Vector{k}^\prime\right)
\end{equation}
that enter when products of fields are Fourier transformed. A tilde
``$~\tilde{ }~$'' always denotes the Fourier transform of a random field or
function beneath the tilde.

We restrict ourselves to the case of \emph{strictly Gaussian fields},
which is a reasonable assumption on linear scales (see e.g. Bernardeau
et al. 2002).  As a consequence, all connected higher order
correlation terms like bispectra vanish, which makes the following
equations a lot simpler.  Further, in the cosmological context the
density fields are \emph{isotropic and homogeneous} random fields.

The correlation power spectrum $P\left(\Vector{k}\right)$ between two
homogeneous random field with the Fourier coefficients
$\tilde{\delta}_1\left(\Vector{k}\right)$ and
$\tilde{\delta}_2\left(\Vector{k}\right)$ respectively is
\begin{equation}
 \left(2\pi\right)^3\delta_D\left(\Vector{k}-\Vector{k}^\prime\right)
 P\left(\Vector{k}\right)\equiv
 \ave{
  \tilde{\delta}_1\left(\Vector{k}\right)
  \tilde{\delta}_2^\ast\left(\Vector{k}^\prime\right)
}
\; ,
\end{equation}
stating that homogeneity requires only the Fourier coefficients of the
same $\Vector{k}$ to be correlated. This relation also states that the
power spectrum $P\left(\Vector{k}\right)$ is related to the correlator
in the following way
\begin{equation}
  P\left(\Vector{k}\right)=
  \int\frac{d^3\Vector{k}^\prime}{\left(2\pi\right)^3}
  \ave{
    \tilde{\delta}_1\left(\Vector{k}\right)
    \tilde{\delta}_2^\ast\left(\Vector{k}^\prime\right)
  } 
\; .
\end{equation}  
Due to this relation, we are going to use a slightly different
definition $\ave{...}^\prime$ of the ensemble average:
\begin{equation}\label{defaveprime}
  \Ave{\tilde{\delta}_1\left(\Vector{k}\right)
    \tilde{\delta}_2^\ast\left(\Vector{k}^\prime\right)}
  \equiv 
  \int\frac{d^3\Vector{k}^\prime}{\left(2\pi\right)^3}
  \ave{\tilde{\delta}_1\left(\Vector{k}\right)
    \tilde{\delta}_2^\ast\left(\Vector{k}^\prime\right)}
  \; ,
\end{equation}
which in the following is useful to derive the differential equations
for the correlation power spectra.  We also introduce the convention
to omit the $\Vector{k}$ arguments for the correlators and the power
spectra. Instead, we use the following notation: Power spectra have as
well as the first field in the two-point correlator (in the above definition
$\tilde{\delta}_1$) as argument always $\Vector{k}$, whereas
the second field in the correlator has the argument
$\Vector{k}^\prime$.  For example, according to this convention the
following two lines are identical:
\begin{eqnarray}
  P\left(\Vector{k}\right)&=&
  \Ave{\tilde{\delta}_1\left(\Vector{k}\right)
    \tilde{\delta}_2\left(\Vector{k}^\prime\right)^\ast}
  \\\nonumber
  P&=&\Ave{\tilde{\delta}_1\tilde{\delta}_2^\ast}
  \;. 
\end{eqnarray}

After explaining the notation, we now accordingly define the
correlation power spectra between the model random fields by
\begin{equation}\label{powerdef}
  P_{ij}=P_{ji}\equiv\Ave{\tilde{\delta}_i\tilde{\delta}_j^\ast}
  \;\;\;\;
  P_i\equiv\Ave{\tilde{\delta}_i\tilde{\delta}_{\rm m}^\ast}
  \;\;\;\;
  P_{\rm m}\equiv\Ave{\tilde{\delta}_{\rm m}\tilde{\delta}_{\rm m}^\ast}
  \; ,
\end{equation}
where $P_{ij}$ is the correlation power spectrum between galaxy
population $n_i$ and $n_j$, thus for $i=j$ the auto-correlation of
population $n_i$. $P_i$ denotes the cross-correlation between the
population $n_i$ and the dark matter field $\rho_{\rm m}$. $P_{\rm m}$
is the dark matter auto-correlation.

To work out their evolution, we first multiply both sides of Eq.
\Ref{basiceqfourier} by $\tilde{\delta}_{\rm
  m}^\ast\left(\Vector{k}^\prime\right)$, take the (modified) ensemble
average $\ave{...}^\prime$ and use the definition of the power spectra
to get
\begin{eqnarray}\label{didm1}
 \Ave{\frac{\partial\tilde{\delta}_i}{\partial
 t}\tilde{\delta}_{\rm m}^\ast}
&=&
 \Ave{\frac{\partial\tilde{\delta}_{\rm m}}{\partial
 t}\tilde{\delta}_{\rm m}^\ast}+\frac{1}{\bar{n}_i}
\left[
\Ave{\tilde{\Phi}_i\tilde{\delta}_{\rm m}^\ast}-
P_i\frac{\partial\bar{n}_i}{\partial t}
\right]
\; .
\end{eqnarray}
Note that all terms containing bispectra (three-point correlations)
have been neglected. They turn up when the correlation of two
convolved fields with a third other field is calculated (see Appendix
A) as for the velocity term in Eq. \Ref{basiceqfourier}.

The equation simplifies further, if we use the following two
relations, obtained by taking the time derivative of the power spectra
definitions \Ref{powerdef}
\begin{eqnarray}\label{pdm}
  \frac{\partial P_{\rm m}}{\partial t}&=&
  \Ave{\frac{\partial\tilde{\delta}_{\rm m}}{\partial
  t}\tilde{\delta}_{\rm m}^\ast}+
  \Ave{\tilde{\delta}_{\rm m}\frac{\partial\tilde{\delta}_{\rm m}^\ast}{\partial
  t}}
  =2 \Ave{\frac{\partial\tilde{\delta}_{\rm m}}{\partial
    t}\tilde{\delta}_{\rm m}^\ast}
\\\label{pidm}
\Ave{\frac{\partial{\tilde{\delta}}_i}{\partial
  t}\tilde{\delta}_{\rm m}^\ast}&=&
\frac{\partial P_i}{\partial t}-
\Ave{\frac{\partial{\tilde{\delta}}_{\rm m}^\ast}{\partial
  t}\tilde{\delta}_i}
\; .
\end{eqnarray}
Eq. \Ref{pdm} utilises the fact that the power spectra are real number
functions, thus identical to its complex conjugate.  Eq.  \Ref{didm1}
can according to Eq. \Ref{pdm} and \Ref{pidm} be written as
\begin{eqnarray}\label{didm2}
&&  \frac{\partial P_i}{\partial t}=
\\\nonumber
&&
  \frac{1}{2}\frac{\partial P_{\rm m}}{\partial t}
  +\frac{1}{\bar{n}_i}
  \left[
    \Ave{\tilde{\Phi}_i\tilde{\delta}_{\rm m}^\ast}-
    P_i\frac{\partial\bar{n}_i}{\partial t}
  \right]
  +  \Ave{\frac{\partial{\tilde{\delta}}_{\rm m}^\ast}{\partial
  t}\tilde{\delta}_i}
\end{eqnarray}
leaving us with an equation for the dark matter-galaxy power spectrum.

As a second step, we try to do a similar thing for the galaxy-galaxy
power spectra $P_{ij}$. Multiplying both sides of \Ref{basiceqfourier}
by $\tilde{\delta}_j^\ast\left(\Vector{k}^\prime\right)$ and taking
the ensemble average yields:
\begin{eqnarray}\label{didj}
 \Ave{\frac{\partial\tilde{\delta}_i}{\partial
 t}\tilde{\delta}_j^\ast}
&=&
 \Ave{\frac{\partial\tilde{\delta}_{\rm m}}{\partial
 t}\tilde{\delta}_j^\ast}+\frac{1}{\bar{n}_i}
\left[
\Ave{\tilde{\Phi}_i\tilde{\delta}_j^\ast}-
P_{ij}\frac{\partial\bar{n}_i}{\partial t}
\right]
\; .
\end{eqnarray}
This is already the first term out of two we need for the time
evolution of $P_{ij}$:
\begin{equation}
  \frac{\partial P_{ij}}{\partial t}
  = 
  \Ave{\frac{\partial\tilde{\delta}_i^\ast}{\partial
      t}\tilde{\delta}_j}+
  \Ave{\frac{\partial\tilde{\delta}_j}{\partial
      t}\tilde{\delta}_i^\ast}
 \; .
\end{equation}
The second is obtained by swapping the indices $i$ and $j$ and taking
the complex conjugate of Eqs. \Ref{didj}.  Combining these eventually
gives
\begin{eqnarray}\label{didj2}
\frac{\partial P_{ij}}{\partial t}
&=&
 \Ave{\frac{\partial\tilde{\delta}_{\rm m}}{\partial
 t}\tilde{\delta}_j^\ast}+\frac{1}{\bar{n}_i}
\left[
\Ave{\tilde{\Phi}_i\tilde{\delta}_j^\ast}-
P_{ij}\frac{\partial\bar{n}_i}{\partial t}
\right]
\\\nonumber
&+&
 \Ave{\frac{\partial\tilde{\delta}_{\rm m}^\ast}{\partial
 t}\tilde{\delta}_i}+\frac{1}{\bar{n}_j}
\left[
\Ave{\tilde{\Phi}_j^\ast\tilde{\delta}_i}-
P_{ij}\frac{\partial\bar{n}_j}{\partial t}
\right]
\; .
\end{eqnarray}

For the next step, we would like to approximate in Eqs.  \Ref{didm2}
and \Ref{didj2} the time derivative $\dt{\tilde{\delta}_{\rm m}}$
using perturbation theory.  For our purposes, the lowest order
approximation of $\tilde{\delta}_{\rm m}$ is sufficient, because we
have restricted the model to large (linear) scales where the
cosmological fields may be considered as Gaussian random fields.
Considering only the growing mode, to lowest order the density field
of the dark matter $\tilde{\delta}_{\rm m}|_{t_i}$ is from one initial
time $t_i$ onwards growing linearly, $\Vector{k}$-independently with
time (e.g.  Bernardeau et al. 2002)
\begin{eqnarray}\nonumber
  \tilde{\delta}_{\rm m}&=&
  D_+~\tilde{\delta}_{\rm m}|_{t_i}
  \\\label{zerothorder}
  \frac{\partial\tilde{\delta}_{\rm m}}{\partial t}&=&
  \frac{\partial D_+}{\partial t}~\tilde{\delta}_{\rm m}|_{t_i}
  =
  \frac{1}{D_+}\frac{\partial D_+}{\partial t}~\tilde{\delta}_{\rm m}
  =
  \frac{\partial \ln{D_+}}{\partial t}~\tilde{\delta}_{\rm m}
\end{eqnarray}
where $D_+$ can be shown to be the integral (e.g. Peacock 1999)
\begin{equation}\label{defdplus}
  D_+\left(a\right)\propto H\left(a\right)
  \int_0^a
  da^\prime\frac{1}{\left[a^\prime H\left(a^\prime\right)\right]^3}
  \; .
\end{equation}
A very good approximation to this integral is given by fitting formula
of Carroll et al. (1992).

Employing the lowest order approximation of $\tilde{\delta}_{\rm m}$
yields for the terms in question
\begin{eqnarray}\nonumber
  \Ave{\frac{\partial\tilde{\delta}_{\rm m}^\ast}{\partial
    t}\tilde{\delta}_i}&=&
  \Ave{\frac{\partial\ln{D_+}}{\partial t}\tilde{\delta}_{\rm m}^\ast\tilde{\delta}_i}=
  \frac{\partial\ln{D_+}}{\partial t}~P_i
  \equiv\frac{R\left(t\right)}{2}~P_i
  \\\nonumber
  \\\nonumber
  \frac{\partial P_{\rm m}}{\partial t}&=&
  \Ave{\frac{\partial\ln{D_+}}{\partial t}\tilde{\delta}_{\rm m}^\ast\tilde{\delta}_{\rm m}}+
  \Ave{\tilde{\delta}_{\rm m}^\ast\frac{\partial\ln{D_+}}{\partial t}\tilde{\delta}_{\rm m}}
  \\\label{linpevolve0}
  &=&
  2\frac{\partial\ln{D_+}}{\partial t}~P_{\rm m}
  \equiv R\left(t\right)~P_{\rm m}
  \; .
\end{eqnarray}
The newly introduced function
\begin{equation}\label{defR}
  R\left(t\right)\equiv 
  \frac{1}{\pow{m}}\dt{\pow{m}}=2\frac{\partial\ln{D_+}}{\partial t}  
\end{equation}
is the rate at which the power spectrum of the dark matter is growing
on linear scales.

Plugging this expression into Eqs.  \Ref{didm2} and \Ref{didj2}
 enables us to write the differential equations
for the correlation power spectra in a closed form
\begin{eqnarray}\label{linpevolve1}
\frac{\partial P_i}{\partial t}-
R\left(t\right)\frac{P_{\rm m}+P_i}{2}&=&
\frac{\Ave{\tilde{\Phi}_i\tilde{\delta}_{\rm m}^\ast}}{\bar{n}_i}-
P_i\frac{\ave{\Phi_i}}{\bar{n}_i}
\\\nonumber
\\\label{linpevolve2}
\frac{\partial P_{ij}}{\partial t}-
R\left(t\right)\frac{P_j+P_i}{2}&=&
\frac{\Ave{\tilde{\Phi}_i\tilde{\delta}_j^\ast}}{\bar{n}_i}+
\frac{\Ave{\tilde{\Phi}_j^\ast\tilde{\delta}_i}}{\bar{n}_j}
\\\nonumber
&-&
P_{ij}\left[
  \frac{\ave{\Phi_i}}{\bar{n}_i}+
  \frac{\ave{\Phi_j}}{\bar{n}_j}
\right] 
\; .
\end{eqnarray}
The terms on the left hand side (lhs) containing $R\left(t\right)$ are
responsible for driving a biased galaxy distribution towards the dark
matter distribution. Setting these terms to zero, switches
off the coupling to the dark matter field due to the bulk flow
assumption.

\subsection{Linear scale evolution of stochastic linear bias}

We define the bias parameters with one index, thus
$r_i$ and $b_i$, to be the bias of the $i$th galaxy population with
respect to the dark matter, whereas two indices, $b_{ij}$ and
$r_{ij}$, denote the bias between the $i$th and $j$th galaxy population:
\begin{eqnarray}\label{linbias}
    b_i\equiv\sqrt{\frac{P_{ii}}{P_{\rm m}}}&&
    r_i\equiv\frac{P_i}{\sqrt{P_{ii}P_{\rm m}}}
    \\\nonumber
    b_{ij}\equiv\sqrt{\frac{P_{ii}}{P_{jj}}}&&
    r_{ij}\equiv\frac{P_{ij}}{\sqrt{P_{ii}P_{jj}}}
    \; .
\end{eqnarray}

Using this definition, we can write down differential equations for
$\left(b_i~r_i~b_{ij}~r_{ij}\right)$ based on Eqs.  \Ref{linpevolve1}
and \Ref{linpevolve2}.  Appendix B shows how this is done in detail.
The main result there is the following set of equations (the equation
for the mean density $\bar{n}_i$ has been added for the sake of
completeness) showing the evolution of the bias parameters for any
kind of interaction term $\Phi_i$:
\begin{eqnarray}\label{bi}
  \dt{b_i}&=&
  R\left(t\right)\frac{r_i-b_i}{2}+I^1_i
  \\\nonumber\\\label{bij}
  \dt{b_{ij}}&=&
  R\left(t\right)
  \frac{r_ib_j-r_jb_i}{2b_ib_j}b_{ij}
  +b_{ij}\left[\frac{I^1_i}{b_i}-\frac{I^1_j}{b_j}\right]
  \\\nonumber\\\label{ri}
  \dt{r_i}&=&
  R\left(t\right)\frac{1-r_i^2}{2b_i}+I^{2}_i
   \\\nonumber\\\label{rij}
  \dt{r_{ij}}&=&
  R\left(t\right)
  \left[\frac{r_i-r_{ij}r_j}{2b_j}+\frac{r_j-r_{ij}r_i}{2b_i}\right]
  +I^{3}_{ij}+\left[I^{3}_{ji}\right]^\ast
  \\\label{linden}
  \dt{\bar{n}_i}&=&
  \ave{\Phi_i}=\left.\ave{\tilde{\Phi}_i}\right|_{k=0}
  \\\nonumber\\\nonumber
  I^0_i&\equiv&
  \frac{1}{\bar{n}_i}\frac{1}{b_i}
    \frac{\Ave{\tilde{\Phi_i}\tilde{\delta}_i^\ast}}{\pow{m}}
  \\\nonumber
  I^1_i&\equiv&
  I^0_i-\frac{b_i}{\bar{n}_i}\ave{\Phi_i}=
  I^0_i-\frac{b_i}{\bar{n}_i}\left.\ave{\tilde{\Phi}_i}\right|_{k=0}
  \\\nonumber
  I^{2}_i&\equiv&\frac{1}{\bar{n}_i}\left[
    \frac{\Ave{\tilde{\Phi}_i\tilde{\delta}_m^\ast}}{\pow{m}}\frac{1}{b_i}-
    \frac{\Ave{\tilde{\Phi}_i\tilde{\delta}_i^\ast}}{\pow{m}}\frac{r_i}{b_i^2}
  \right]
  \\\label{interactcorr}
  I^{3}_{ij}&\equiv&
  \frac{1}{\bar{n}_i}\left[
    \frac{\Ave{\tilde{\Phi}_i\tilde{\delta}_j^\ast}}{\pow{m}}\frac{1}{b_ib_j}-
    \frac{\Ave{\tilde{\Phi}_i\tilde{\delta}_i^\ast}}{\pow{m}}\frac{r_{ij}}{b_i^2}
  \right]
  \; .
\end{eqnarray}
The interaction terms $I^1_i$, $I^{2}_i$ and $I^{3}_{ij}$ vanish for
$\Phi_i=0$; they are responsible for deviations from the
interaction-free evolution of the linear bias parameters. In Eq.
\Ref{linden}, we equivalently expressed the interaction rate in terms
of the Fourier representation of $\Phi_i$. Depending on the definition
of $\Phi_i$ this representation can be mathematically of advantage,
especially when derivatives or integrals are involved.

 \begin{figure*}
   \begin{center}
     \epsfig{file=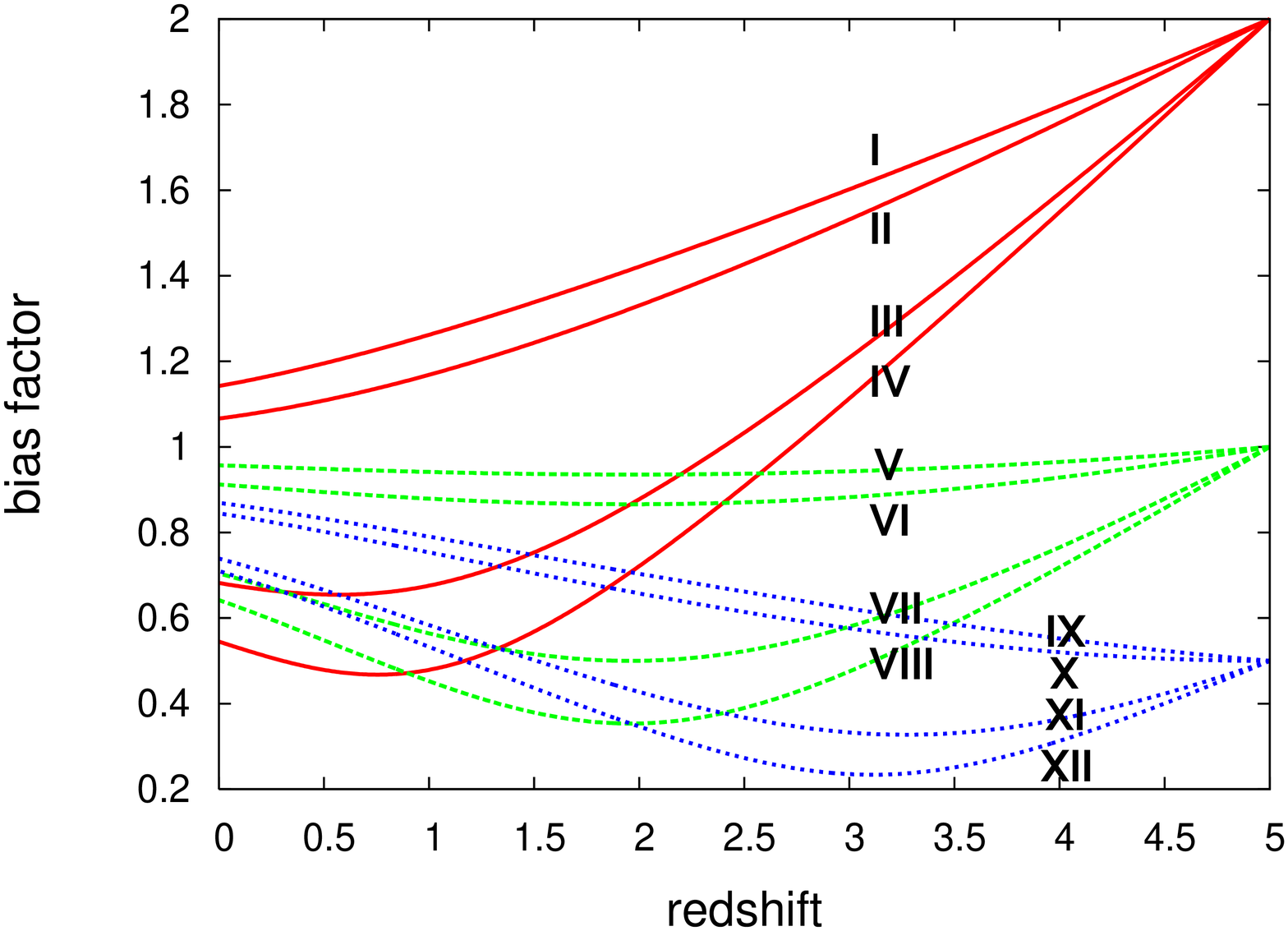,width=87mm,angle=0}
     \epsfig{file=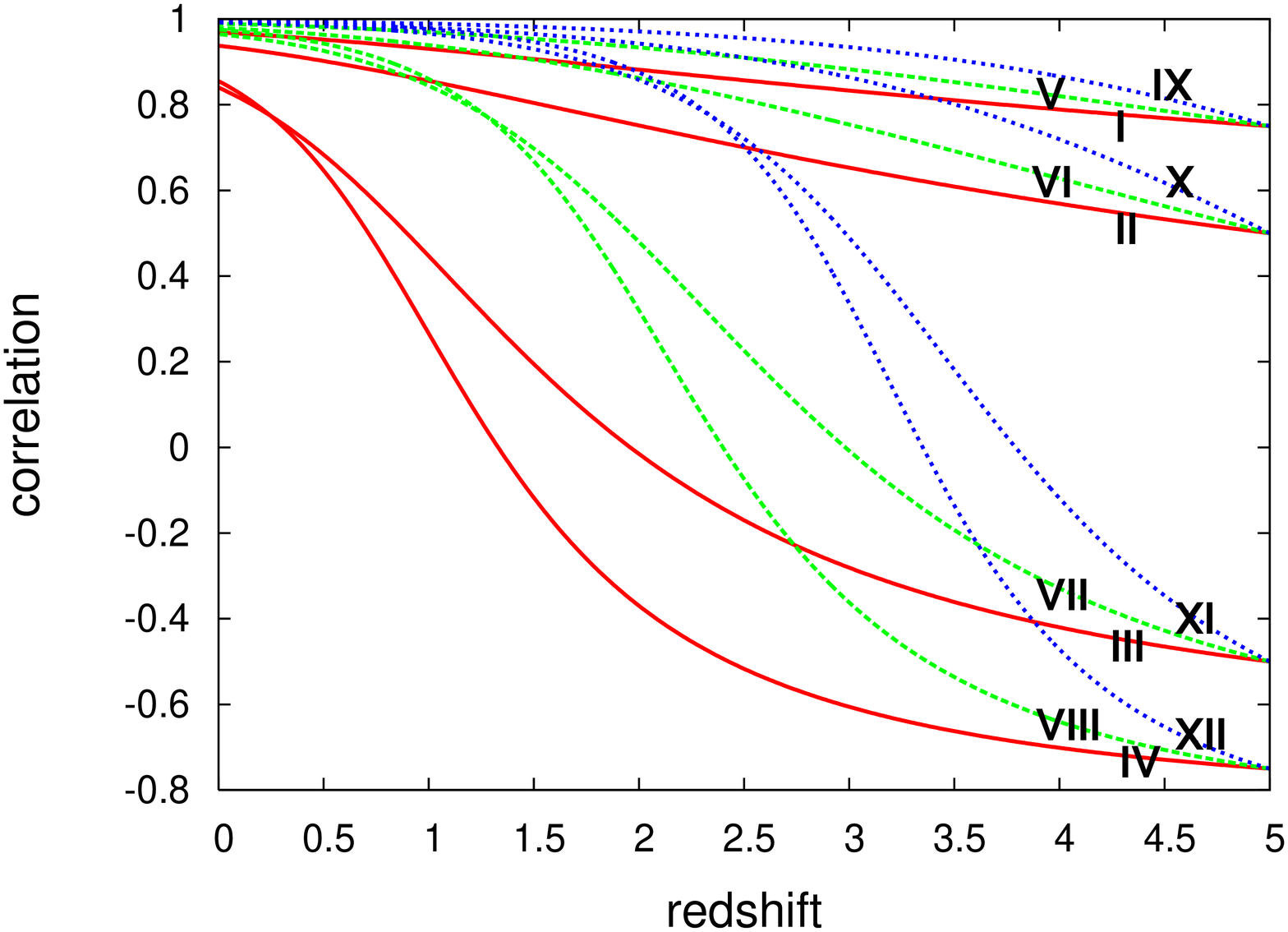,width=87mm,angle=0}
   \end{center}
   \caption{Evolution of the linear bias with no coupling between
     the galaxy species or to the dark matter present; the number of
     galaxies is hence conserved. One curve from the right and one
     curve from the left panel always belong together for one plotted
     model, twelve models are presented (roman numbers). The left
     panel shows the bias $b$ evolving for three quadruple of models
     from the initial values $b=2, 1, 0.5$ at redshift $z=5$ to $z=0$;
     the curves of each quadruple belong to initially (from upper to
     lower): $r=0.75, 0.5, -0.5, -0.75$. In the right panel we depict
     the corresponding correlation parameter.}
   \label{noninteractbias}
   \begin{center}
     \epsfig{file=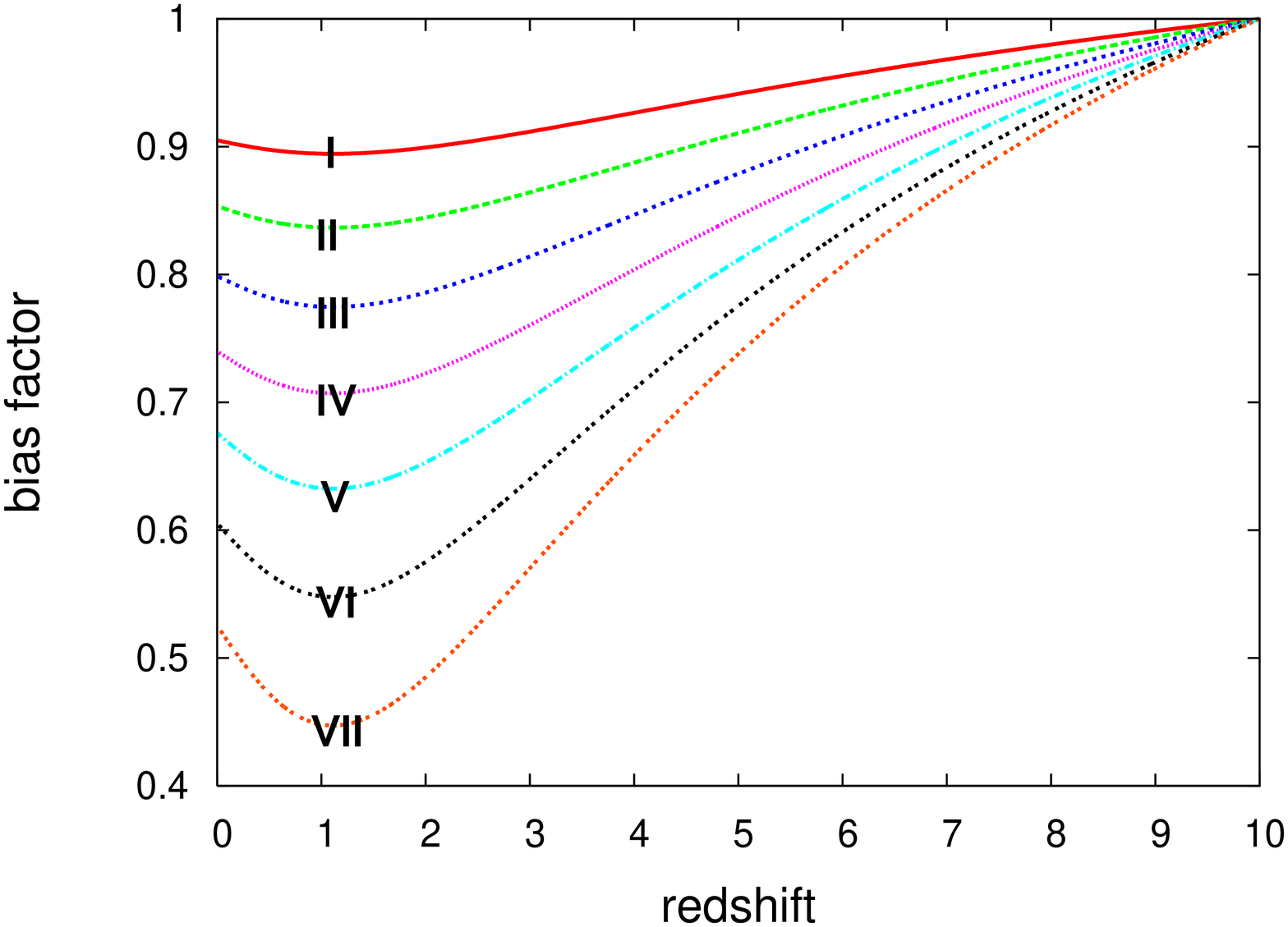,width=87mm,angle=0}
     \epsfig{file=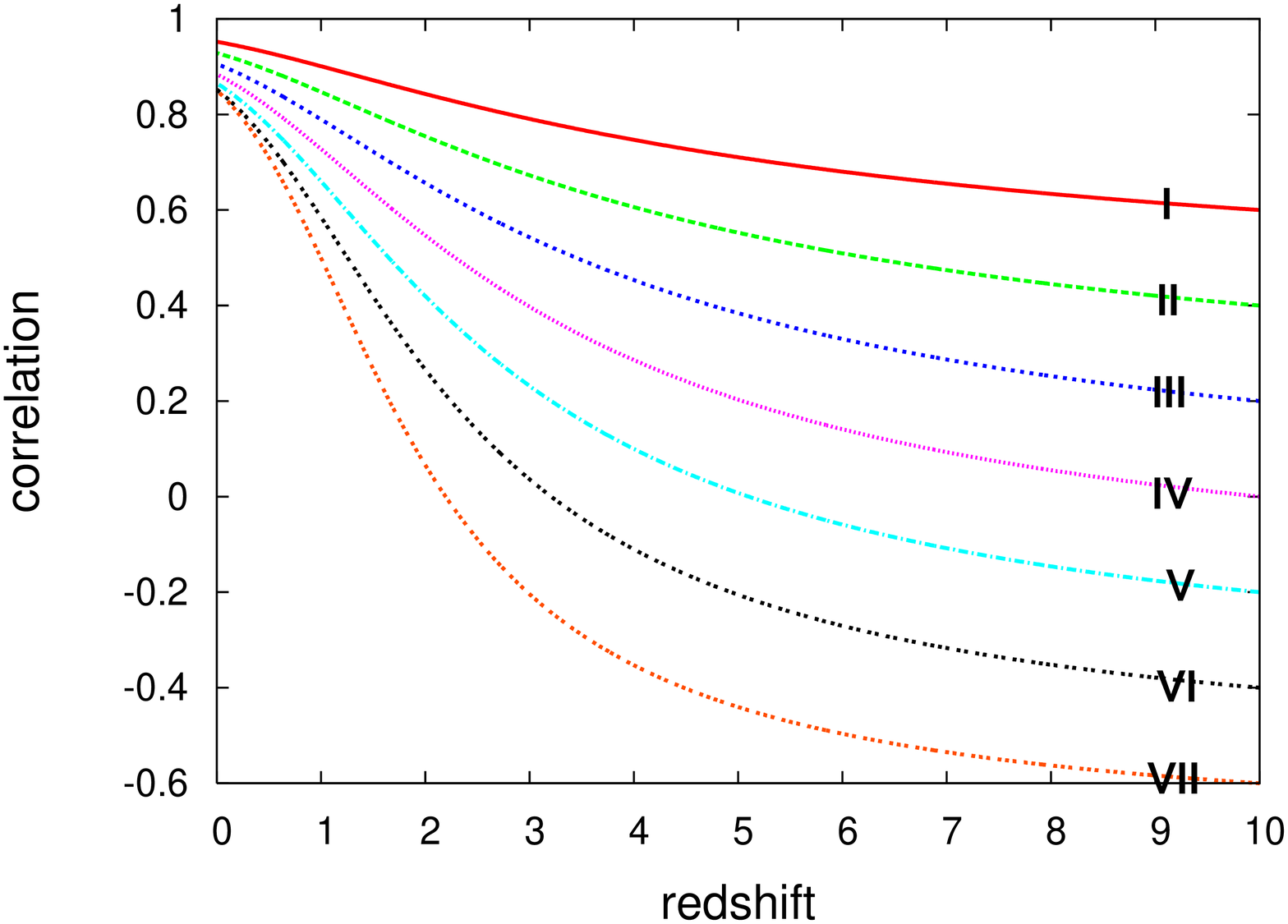,width=87mm,angle=0}
   \end{center}
   \caption{Evolution of the relative linear bias between two galaxy species
     , both starting off at $z=10$ with $b_1=b_2=4$.  The correlation
     of one species to the dark matter is always $r_1=1$, whereas the
     second species has $r_2=0.6,0.4,0.2,0.0,-0.2,-0.4,-0.6$ for the
     curves in the left panel (upper to lower). The initial
     correlations \emph{between} the galaxies where chosen to be
     $r_{12}=0.6,0.4,0.2,0.0,-0.2,-0.4,-0.6$. The left panel plots the
     evolution of $b_{12}$, the right panel $r_{12}$, same roman
     numbers correspond to one model. No coupling is present, hence
     the galaxy number is conserved.}
   \label{noninteractrelbias}
 \end{figure*}

\subsection{Bias evolution without interaction}
 
With no interaction present, $\Phi_i=0$, the model treats the same
case as in the second section of TP98.  Fig.  \ref{noninteractbias}
shows a diagram similar to the one in their paper: it can be seen that
an initially biased galaxy distribution is more and more relaxing
towards the dark matter distribution, asymptotically closing in to
$r=1$ and $b=1$ (``debiasing''). That this is indeed a stationary
state, i.e.  $\dt{b_i}=\dt{r_i}=0$, can be seen from Eqs. \Ref{bi} and
\Ref{ri} for which the only stationary solutions are $b_i=r_i=\pm 1$
(without interaction, hence $I^1_i=0$).

The second solution with $b_i=r_i=-1$ has to be excluded, because the
bias factor is by definition always positive. The only possible way to be
attracted by this stationary point is that we have $r_i=-1$ at all
time. For all other values $r_i>-1$, the correlation parameter is an
increasing function with time, inevitably approaching the other
stationary solution. This peculiarity is therefore avoided if we
exclude $r_i=-1$ as initial condition.
 
The bias \emph{between} two galaxy populations also has a stationary
solution at $b_{ij}=b_i/b_j=r_{ij}=1$. This follows from Eqs.
\Ref{bij} and \Ref{rij} ($I^2_i=I^3_{ij}=0$).  Fig.
\ref{noninteractrelbias} shows as an illustration the evolution of the
relative bias between two galaxy populations while they are getting
debiased with respect to the dark matter.

\section{Linear and quadratic interaction rates $\Phi_i$}

\begin{table*}
\parbox{9cm}{\parbox{8cm}{
\caption{\label{tab}
Three tables listing the contributions of the different couplings
in Sect. 3 to the interaction terms $I^0_i$, $I^2_i$, $I^3_{ij}$ 
and the mean interaction rate $\ave{\Phi_i}$ sorted by the coupling
constants; they are required by  Eqs. \Ref{bi}
to \Ref{linden}. $A_i$ corresponds to a constant galaxy
production/destruction, $B_i^l$ couples galaxy field $n_l$ to $n_i$ (linear),
$C_i$ $n_i$ to the dark matter field $\rho_{\rm m}$ (linear), $D_i^{ls}$
couples $n_l$ 
and $n_s$ to $n_i$ (quadratic), $E_i$ couples $\rho_{\rm m}^2$
to $n_i$ (quadratic), and $F_i^l$ couples 
$n_i$ to $n_l$ and the dark matter field $\rho_{\rm m}$ (quadratic). 
The whole expression contributing is the product between the coupling
constant, first column, and the expression in the second column or
third column.
Different contributions from different couplings are just added; \textbf{we
are using Einstein's summation convention for the variables $l$ and $s$}.
Note that we have the special case $r_{ii}\equiv 1$ 
by definition of the correlation
parameter. 
The bias parameters $\widehat{b_ir_i}$ and $\widehat{b_jb_jr_{ij}}$,
and $\ave{\delta_{\rm m}^2}$ are explained in section
3. They are only needed for modelling the mean galaxy density in the
presence of quadratic couplings.}}}
\parbox{9cm}{\begin{tabular}{lll}
  \hline\\      
  \textbf{term}
  & 
  $\bf I^0_i$
  &
  $\bf\ave{\Phi_i}$
  \\\\\hline\\
  $\bf A_i$ &
  $0$ &
  $1$
  \\\\
  $\bf B_i^l$ &
  $\frac{\bar{n}_l}{\bar{n}_i}r_{il}b_l$ &
  $\bar{n}_l$
  \\\\
  $\bf C_i$ &
  $\frac{1}{\bar{n}_i}r_i$ &
  $1$
  \\\\
  $\bf D_i^{ls}$&
  $\frac{\bar{n}_l\bar{n}_s}{\bar{n}_i}
  r_{li}b_l+r_{si}b_s$ &
  $\left(1+\widehat{b_lb_sr_{ls}}\ave{\delta_{\rm m}^2}\right)
  \bar{n}_l\bar{n}_s$
  \\\\
  $\bf E_i$&
  $\frac{2}{\bar{{n}}_i}r_i$ &
  $1+\ave{\delta_m^2}$
  \\\\
  $\bf F_i^l$ &
  $\frac{\bar{{n}}_l}{\bar{{n}}_i}
  \left(r_i+r_{li}b_l\right)$ &
  $\left(1+\widehat{b_lr_l}\ave{\delta_{\rm m}^2}\right)
  \bar{n}_l$
  \\\\\hline\\
\end{tabular}}
\parbox{9cm}{\begin{tabular}{ll}
    \hline\\      
    \textbf{term}
    & 
    $\bf I^2_i$
  \\\\\hline\\
  $\bf A_i$ & 0
  \\\\
  $\bf B_i^l$ &
  $\frac{\bar{{n}}_l}{\bar{{n}}_i}\frac{b_l}{b_i}\left(r_l-r_ir_{li}\right)$
  \\\\
  $\bf C_i$ &
  $\frac{1}{\bar{{n}}_i}\frac{1}{b_i}\left(1-r_i^2\right)$
  \\\\
  $\bf D_i^{ls}$&
  $\frac{\bar{{n}}_l\bar{{n}}_s}{\bar{{n}}_i}\frac{1}{b_i}
  \left(\left[r_l-r_{li}r_i\right]b_r+\left[r_s-r_{si}r_i\right]b_s\right)$
  \\\\
  $\bf E_i$&
  $\frac{2}{\bar{{n}}_i}\frac{1}{b_i}\left(1-r_i^2\right)$ 
  \\\\
  $\bf F_i^l$ &
  $\frac{\bar{{n}}_l}{\bar{{n}}_i}\frac{1}{b_i}
  \left(1-r_i+\left[r_l-r_{li}\right]b_l\right)$
  \\\\\hline\\
\end{tabular}}
\parbox{9cm}{\begin{tabular}{ll}
 \hline\\        
 \textbf{term}
  & 
  $\bf I^3_{ij}$
  \\
  \\\hline\\
  $\bf A_i$ & 0
  \\\\
  $\bf B_i^l$ &
  $\frac{\bar{{n}}_l}{\bar{{n}}_i}\frac{b_l}{b_i}\left(r_{lj}-r_{li}r_{ij}\right)$
  \\\\
  $\bf C_i$ &
  $\frac{1}{\bar{{n}}_i}\frac{1}{b_i}\left(r_j-r_ir_{ij}\right)$
  \\\\
  $\bf D_i^{ls}$&
  $\frac{\bar{{n}}_l\bar{{n}}_s}{\bar{{n}}_i}\frac{1}{b_i}   
  \left(\left[r_{lj}-r_{ij}r_{li}\right]b_l+\left[r_{sj}-r_{ij}r_{si}\right]b_s\right)$
  \\\\
  $\bf E_i$&
  $\frac{2}{\bar{{n}}_i}\frac{1}{b_i}\left(r_j-r_ir_{ij}\right)$
  \\\\
  $\bf F_i^l$ &
  $\frac{\bar{{n}}_l}{\bar{{n}}_i}\frac{1}{b_i}
  \left(\left[r_{lj}-r_{li}r_{ij}\right]b_l+r_j-r_ir_{ij}\right)$
  \\\\\hline\\
\end{tabular}}
\end{table*}
 
To be specific about the interaction term, we make the following
 Ansatz for $\Phi_i$, namely a Taylor expansion in $n_i$ and 
$\rho_{\rm m}$ up to second order:
\begin{equation}\label{phidef}
  \Phi_i=A_i+B_i^rn_r+\hat{C}_i\rho_{\rm m}
  +D_i^{rs}n_rn_s+\hat{E}_i\rho_{\rm
  m}^2+\hat{F}_i^r\rho_{\rm m}n_r
  \; .
\end{equation}
$A_i$, $B^r_i$, $\hat{C}_i$, $D^{rs}_i$, $\hat{E}_i$ and $\hat{F}_i^r$
are phenomenological coupling constants. Note that we are using the
Einstein summing convention that abbreviates e.g. the expression
$\sum_{rs}D^{rs}_in_rn_s$ through $D^{rs}_in_rn_s$. As before, we skip
the position arguments of the density fields.

This particular $\Phi_i$ is motivated by the idea that locally the
galaxy density may be changed - apart from converging or diverging
bulk flows - by galaxy collisions or mergers with interaction rates
proportional to the product of the density fields involved
($D_i^{rs}$, $\hat{E}_i$ and $\hat{F_i^r}$).  In addition, we also
include all lower order terms, like, for instance, a constant rate of
galaxy production $A_i$ or a rate that is linear with some density
field ($B_i^r$ and $\hat{C}_i$). As the non-linear, quadratic
couplings linear couplings may also have a physical 
interpretation in this context: a galaxy of one
population is with a constant probability - independent of its
environment- transformed into a member of another
population (passive evolution).

Actually needed inside Eqs. \Ref{bi} to \Ref{linden} are, however, not
the $\Phi_i$ but the interaction terms in Eqs. \Ref{interactcorr}.
Those are mainly functions of the \emph{interaction correlators}
$\Ave{\tilde{\Phi}_i\tilde{\delta}_j^\ast}$ and
$\Ave{\tilde{\Phi}_i\tilde{\delta}_m^\ast}$ whose evaluation can be
found in Appendix C.

We have to evaluate the interaction rate per unit volume in Eq.
\Ref{linden}, too:
\begin{eqnarray}\label{phiv}
 \ave{\Phi_i}&=&
 A_i+C_i+E_i+\left(B_i^r+F_i^r\right)\bar{n}_r+D_i^{rs}\bar{n}_r\bar{n}_s
 \\\nonumber
 &+&
 D_i^{rs}\bar{n}_r\bar{n}_s\ave{\delta_r\delta_s}+E_i\ave{\delta_{\rm m}^2}+
 F_i^r\bar{n}_r\ave{\delta_{\rm m}\delta_r}
  \;,
\end{eqnarray}
where the mean dark matter density $\bar{\rho}_{\rm m}$ has been
absorbed inside the coupling constants (Appendix C). Note that the
density contrasts here are in real space.  For linear couplings only,
the evolution of the mean volume density of galaxies is apparently
independent from the way the galaxies are clustered, because then Eq.
\Ref{linden} only depends on $\bar{n}_i$.  Quadratic couplings,
however, introduce the terms like $\ave{\delta_i\delta_j}$, so that
the mean density evolution gets linked to the correlations between
$\delta_i$ and $\delta_{\rm m}$, and the fluctuations of these fields.
The meaning of this is, that under quadratic couplings the mean
density of highly clustered galaxies evolves in a different way than a
completely homogeneous galaxy field.

To develop the last equation a bit further, we now would like to express
the (real space) fluctuations/correlations $\ave{\delta_i\delta_j}$
and $\ave{\delta_{\rm m}\delta_j}$ in terms of linear bias parameters
and the dark matter density fluctuations $\ave{\delta_{\rm m}^2}$ only.
Expanding the correlator $\ave{\delta_i\delta_j}$ in Fourier space
employing Eqs. \Ref{linearbias} gives
\begin{eqnarray}\nonumber
  \ave{\delta_i\delta_j}
  &=&\frac{1}{2\pi^2}\int dk~k^2
  \left|\tilde{W}\left(k\right)\right|^2b_i\left(k\right)b_j\left(k\right)r_{ij}\left(k\right)
  P_{\rm m}\left(k\right)
  \label{smallscale}
  \\
  &=& \widehat{b_ib_jr_{ij}}\;\ave{\delta_{\rm m}^2}
  \\\label{deltamsqd}
  \ave{\delta_{\rm m}^2}&=&
    \frac{1}{2\pi^2}\int dk~w\left(k\right)
  \; ,
\end{eqnarray}
using the definitions
\begin{eqnarray}
  \widehat{b_ib_jr_{ij}}&\equiv&
  \frac{\int
    dk~w\left(k\right)b_i\left(k\right)b_j\left(k\right)r_{ij}\left(k\right)}
  {\int dk~w\left(k\right)}
  \\\label{weighting}
  w\left(k\right)&\equiv&
  k^2\left|\tilde{W}\left(k\right)\right|^2P_{\rm m}\left(k\right)
  \;.
\end{eqnarray}
We have introduced a \emph{window function} $\tilde{W}\left(k\right)$
to account for the fact that the density fields $\delta_i$ and
$\delta_{\rm m}$ entering the interaction rate $\Phi_i$ as quadratic
coupling terms in general may be smoothed with some kernel
$W\left(r\right)$. It is, for instance, plausible that fluctuations of
the fields much smaller than the typical size of a galaxy are not
relevant for galaxy interactions, although mathematically the density
fields may have an infinite resolution. In that particular case,
$W\left(r\right)$ could be modelled as a top hat of some typical width
$R_{\rm int}$ with the following $\tilde{W}\left(k\right)$ 
\begin{equation}
  \tilde{W}\left(y\right)=\frac{3}{y^3}\left(\sin{y}-y\cos{y}\right)
  \; ,
\end{equation}
with $y\equiv kR_{\rm int}$ (e.g. Peacock 2001, page 500).

\begin{figure}
  \begin{center}
    \epsfig{file=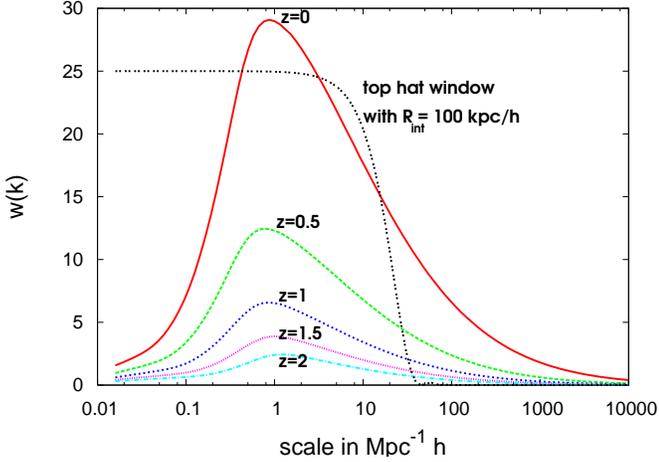,width=90mm,angle=0}
  \end{center}
  \caption{Weighting factors $w\left(k\right)$ (in arbitrary units, see
    Eq. \ref{weighting}) for different redshifts, the window $\tilde{W}\left(k\right)$
    was set to one. The weight maximum stays roughly at the same scale
    in the plotted redshift interval; the peak is quite broad
    however. Also plotted here is the window function 
    $\left|\tilde{W}\left(k\right)\right|^2$ (not normalised) of a top hat
    $W\left(r\right)$ with  $R_{\rm int}=100~\rm kpc/h$ which would
    cut-off the weights beyond $k_{\rm int}/2\pi=10~\rm Mpc^{-1}h$. See Sect. 4 for cosmological
    parameters; the PD96 prescription was used for non-linear regime.}
  \label{weights}
\end{figure}

The expression
$\widehat{b_ib_jr_{ij}}$ is the weighted mean of $b_i\left(k\right)b_j\left(k\right)r_{ij}\left(k\right)$ over
all scales. Fig. \Ref{weights} shows the weights $w\left(k\right)$ for
some redshifts and one particular cosmological model. In the plotted
redshift range, the weight peaks at about $1$ Mpc $h^{-1}$, but has a 
considerable width though; that is assuming that $R_{\rm
  int}\ll 1$ Mpc $h^{-1}$. In an analogue manner, we obtain
\begin{equation}
  \ave{\delta_i\delta_{\rm m}}=
  \widehat{b_ir_i}\;\ave{\delta_{\rm m}^2};\;\;
  \widehat{b_ir_i}\equiv
  \frac{\int
    dk~w\left(k\right)b_i\left(k\right)r_i\left(k\right)}
  {\int dk~w\left(k\right)}
  \; .
\end{equation}
Eq. \Ref{phiv} hence can be written as
\begin{eqnarray}\label{phiv2}
 \ave{\Phi_i}&=&
 A_i+C_i+E_i+
 B_i^r\bar{n}_r
  \\\nonumber
 &+&
 D_i^{rs}\left[1+\widehat{b_rb_sr_{rs}}\ave{\delta_{\rm m}^2}\right]
 \bar{n}_r\bar{n}_s\\\nonumber
 &+&
 F_i^s\left[1+\widehat{b_sr_s}\ave{\delta_{\rm m}^2}\right]\bar{n}_s+
 E_i \ave{\delta_{\rm m}^2}
\; .
\end{eqnarray}

Table \ref{tab} summarises the final result for $I^0_i$, $I^2_i$,
$I^3_{ij}$ and $\ave{\Phi_i}$ as list of contributions stemming from
the various linear and quadratic interaction terms in \Ref{phidef}. As
both the interaction terms and the mean galaxy interaction rate are
linear in $\Phi_i$, all different coupling contributions are simply
added in order to obtain the final terms.

To give an example, assume we would like to couple linearly a galaxy
population $n_1$ to the galaxy population $n_2$; this is an
interaction of the $B_i^j$ type. In our notation,
$\left(b_1~r_1\right)$, $\left(b_2~r_2\right)$ and
$\left(b_{12}~r_{12}\right)$ are the linear bias parameters of
population $n_1$ with respect to the dark matter,
of population $n_2$ with respect to the dark matter and 
of population $n_1$ with respect to
population $n_2$ respectively. According to table \ref{tab}, the
interaction terms are explicitly (after some algebra using
$r_{11}=r_{22}=1$ and $r_{12}=r_{21}$):
\begin{eqnarray}\nonumber
 I^0_1=B_1^1~b_1+\frac{\bar{n}_2}{\bar{n}_1}B_1^2~r_{12}b_2;
 &&
 I^0_2=B_2^2~b_2+\frac{\bar{n}_1}{\bar{n}_2}B_2^1~r_{12}b_1
 \\\nonumber
 I_1^2=B_1^2~\frac{\bar{n}_2b_2}{\bar{n}_1b_1}\left(r_2-r_1r_{12}\right);
 &&
 I_2^2=B_2^1~\frac{\bar{n}_1b_1}{\bar{n}_2b_2}\left(r_1-r_2r_{12}\right)
 \\\nonumber
 I^3_{12}=B_1^2~\frac{\bar{n}_2b_2}{\bar{n}_1b_1}\left(1-r_{12}^2\right);
 &&
 I^3_{21}=B_2^1~\frac{\bar{n}_1b_1}{\bar{n}_2b_2}\left(1-r_{12}^2\right)
 \\\nonumber
 I^3_{11}=0;
 &&
 I^3_{22}=0
 \\\nonumber
 \ave{\Phi_1}=B_1^1\bar{n}_1+B_1^2\bar{n}_2;&&
 \ave{\Phi_2}=B_2^1\bar{n}_1+B_2^2\bar{n}_2
\end{eqnarray}
$B_1^1$ and $B_2^2$ couple the galaxy population to themselves and are
therefore zero if solely couplings between $n_1$ and $n_2$ are
allowed.  If the number of galaxies $\bar{n}_1+\bar{n}_2$ is conserved
by this kind of interaction, thus $\dt{\bar{n}_1}=\dt{\bar{n}_2}=0$,
then we have the further constraint $B_1^2=-B_2^1$ as can be seen from
Eq. \Ref{linden}. In case that two galaxies of $i=1$ ``merge'' to
produce one $i=2$ galaxy, we have the constraint
$B_1^2=-\frac{1}{2}B_2^1$.


\begin{figure}
  \begin{center}
    \epsfig{file=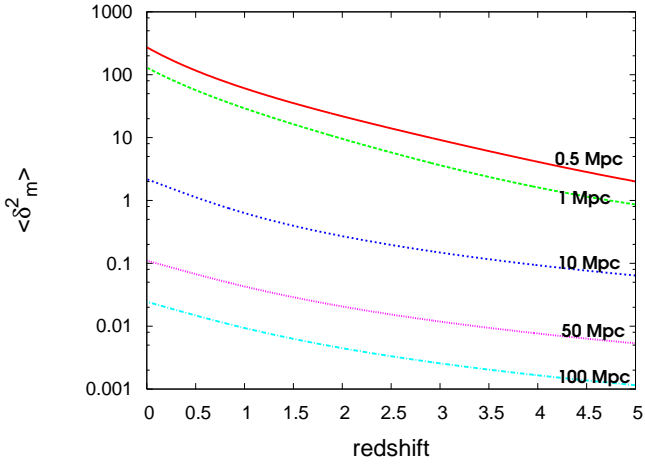,width=90mm,angle=0}
  \end{center}
  \caption{Estimated fluctuations $\ave{\delta_m^2}$, Eq. \Ref{deltamsqd}, of the dark
    matter density field for different cutoffs $R_{\rm int}$ using the
    PD96 prescription for the non-linear clustering regime; see Sect.
    4 for cosmological parameters.}
  \label{dmsqd}
\end{figure}

\section{Toy models}

In this section, we present a few examples to illustrate the impact of
interactions on the evolution of the linear bias parameters. These
include the bias of each galaxy population with respect to both the
dark matter and all other populations.  For predicting the bias
evolution on large scales, we incorporate the model Eqs.  \Ref{bi} to
\Ref{linden}.

Owing to the large number of free parameters and ways to combine them,
there are many models to look at. To explore some of them, we focus on
two galaxy populations and ``switch on'' only one parameter out of
$A_i-F_i^j$ in Eq. \Ref{phidef} while setting the others to zero. This
allows us to look at the effect of the coupling parameters separately.

The evolution is plotted in redshift. Therefore, we have to transform
the derivatives with respect to cosmic time $t$
\begin{equation}
  \frac{\partial}{\partial t}=
  \frac{\partial a}{\partial t}\frac{\partial z}{\partial
  a}\frac{\partial}{\partial z}=
  -\left.\frac{H\left(a\right)}{a}\right|_{a=\left(1+z\right)^ {-1}}
  \frac{\partial}{\partial
  z}
  \; .
\end{equation}
The dark matter growth rate defined in Eq. \Ref{defR} is accordingly
as function of redshift
\begin{equation}
  R\left(z\right)=R\left(t\right)|_{t=t\left(z\right)}=
  \left.2\left[aH\left(a\right)
  \frac{\partial\ln{D_+}}{\partial a}\right]\right|_{a=\left(1+z\right)^{-1}}
 \end{equation}
 which then may be evaluated using Eq. \Ref{defdplus}
 ($H\left(a\right)$ is defined in Eq. \ref{cosmology}).  

 It may be
 useful to have these expressions for a simple cosmology, like for the
 Einstein-de Sitter Universe with $D_+=a$ and
 $H\left(a\right)=H_0a^{-3/2}$ ($a\equiv 1$ at $z=0$)
\begin{equation}
  \left.
  \begin{array}{lll}
    R\left(z\right)&=&+2H_0\left(1+z\right)^{3/2}
    \\\
    \dt{}&=&-H_0\left(1+z\right)^{5/2}\frac{\partial}{\partial z}
  \end{array}
  \right\}
  {\rm Einstein-de~Sitter}
 \; .
\end{equation}

Our cosmology in the examples stated here is a $\Lambda$CDM model with
$\Omega_{\rm m}=0.3$, $\Omega_\Lambda=0.7$, $\rm H_0=70~\rm km
s^{-1}Mpc^{-1}$. Furthermore, a scale-invariant $n=1$ Harrison
Zel'dovich spectrum for the primordial fluctuations is assumed. For
the 3-D power spectrum of the matter fluctuations we use the fitting
formula of Bardeen et al. (1986) for the transfer function, and the
Peacock~\&~Dodds (1996), hereafter PD96, description for the evolution
in the non-linear regime. The power spectrum normalisation is
parameterised with $\sigma_8=0.9$ and the shape parameter assumed to
be $\Gamma=0.21$ (the 3-D matter
fluctuations spectrum is needed for the quadratic coupling models
only).

For the discussion of the toy models see section 5.

\subsection{Constraints on the correlation parameter}

As initial condition, one can set the bias parameter $b_i$ freely. The
relative bias $b_{ij}$ between the different galaxy populations is
thereby also fixed, namely $b_{ij}=b_i/b_j$.

The choice of the initial conditions of the correlation coefficients
$\left(r_i~r_{ij}\right)$ is not free, however.  For example, we
cannot demand population A to be 100 percent correlated to both population B
and population C, but, at the same time, population B to be not correlated
to C. To be more general, we arrange the density contrasts of the dark
matter and $N$ galaxy fields in terms of one single vector
\begin{equation}
  \Vector{x}\left(\Vector{k}\right)=
  \left( \tilde{\delta}_{\rm m}\left(\Vector{k}\right)~~
    \tilde{\delta}_1\left(\Vector{k}\right)~~...~~
    \tilde{\delta}_{\rm N}\left(\Vector{k}\right)\right)^{\rm t}
  \; ,
\end{equation}
with $\Vector{x}^{\rm t}$ being the transpose of $\Vector{x}$.
Concerning the bias parameter, we are restricted by the fact that the
covariance matrix 
$\Matrix{C}\left(\Vector{k}\right)=
\Ave{\Vector{x}\left(\Vector{k}\right)\Vector{x}^{\rm
    t}\left(\Vector{k}^\prime\right)}$ has to be positive
semidefinite, thus the determinant of
\begin{eqnarray}\nonumber
&&  \Matrix{C}\left(\Vector{k}\right)=
\\
&&
  \pow{m}\left(\Vector{k}\right)\left(
    \begin{array}{lllll}
      1 & r_1b_1  & r_2b_2 & ... & r_Nb_N \\
      r_1b_1 & b_1^2 & b_1b_2r_{12} & ... & b_1b_Nr_{1N} \\
      ... & ... & ... & ... & ...\\
      r_Nb_N & b_1b_Nr_{1N} & ... & ...& b_N^2
    \end{array}
  \right)
\end{eqnarray}
has to be greater than or equal to zero (as before, we have left out
the $\Vector{k}$-dependence of the bias parameter in the notation).

For three random fields (or two galaxy populations plus the dark
matter field), this statement is equivalent to
\begin{equation}\label{inequ1}
  2 r_1 r_2 r_{12}\ge r_1^2+r_2^2+r_{12}^2-1
  \; ,
\end{equation}
if the definitions of $\Matrix{C}$, $r_1$, $r_2$, $r_{12}$ are used
(calculation not shown here). It holds for all scales and the
large-scale parameter considered in particular.
Going back to the example above, it
follows immediately from this equation that if we fix two of the three
correlation coefficients with one, say $r_1=r_2=1$, the third
automatically is also forced to be one.  Even more general, if only
one of the correlations is set to one, say $r_1$, then the other two
have to be equal, since we are told by the above constraint that
\begin{equation}
  \left(r_2-r_{12}\right)^2\le 0
  \; .
\end{equation}

Already for four random fields (or three galaxy populations plus
the dark matter field) this condition of positive semi-definiteness
becomes rather lengthy:
\begin{eqnarray}
 && r_1^2r_{23}^2+r_2^2r_{13}^2+r_3^2r_{12}^2+\\\nonumber
 && 2\big[r_1r_2r_{12}+r_1r_3r_{13}+r_2r_3r_{23}+r_{12}r_{13}r_{23}-\\\nonumber
 && r_1r_2r_{13}r_{23}-r_1r_3r_{12}r_{23}-r_2r_3r_{12}r_{13}\big]\\\nonumber
 &&\ge r_1^2+r_2^2+r_3^2+r_{12}^2+r_{13}^2+r_{23}^2-1
  \; .
\end{eqnarray}
By setting all correlations with the third population to zero
($r_3=r_{13}=r_{23}=0$), one can see that this reduces to the forgoing
inequality \Ref{inequ1}. Thus, the constraint for three galaxy
populations is a more general expression that simplifies to the
condition for two populations if one of the three galaxy populations
is not all correlated to the two others and the dark matter; it is in
a statistical sense disconnected from the others.

\begin{figure*}
  \begin{center}
    \begin{tabular}{cc}
    \epsfig{file=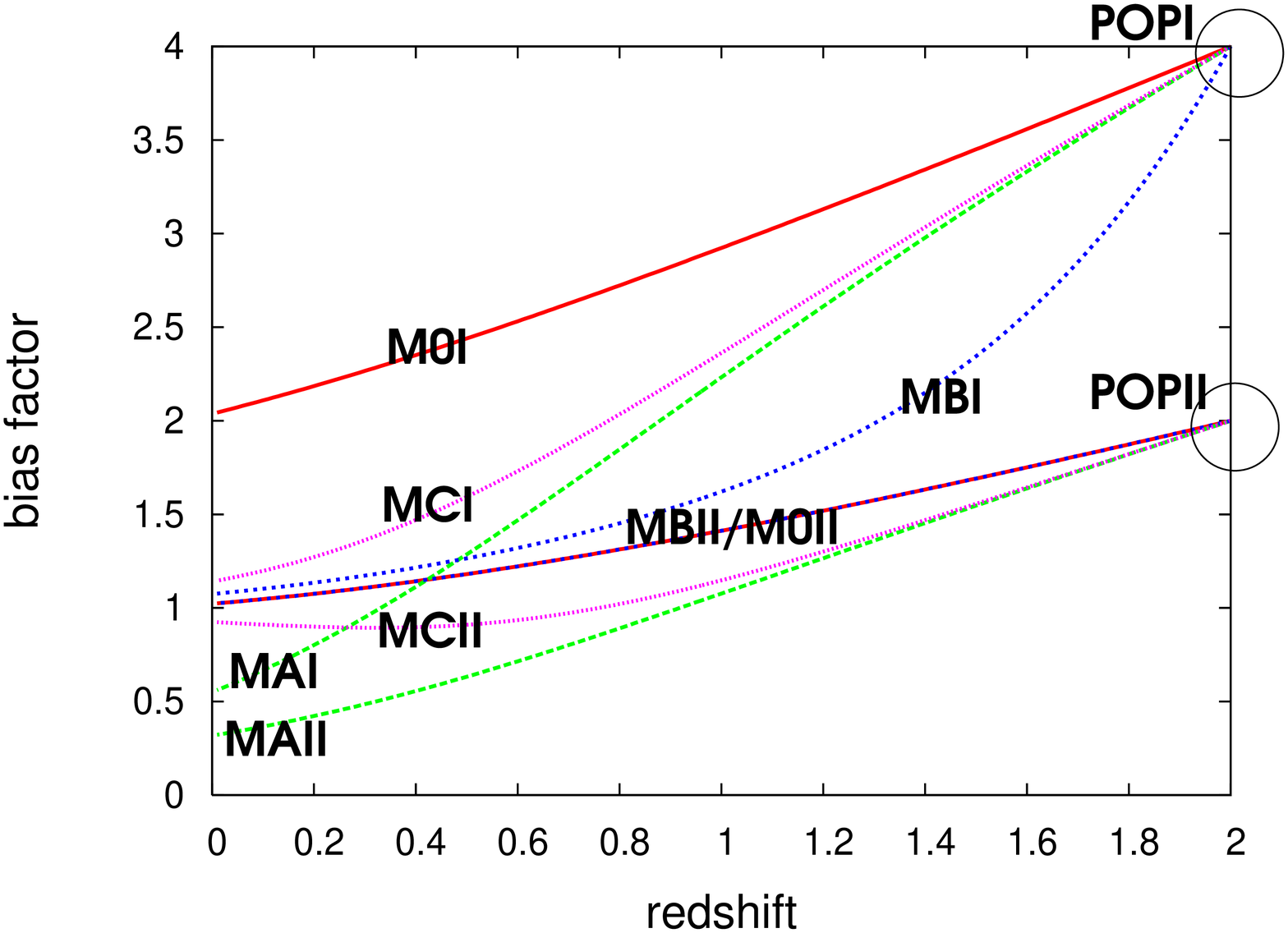,width=80mm,angle=0}&
    \epsfig{file=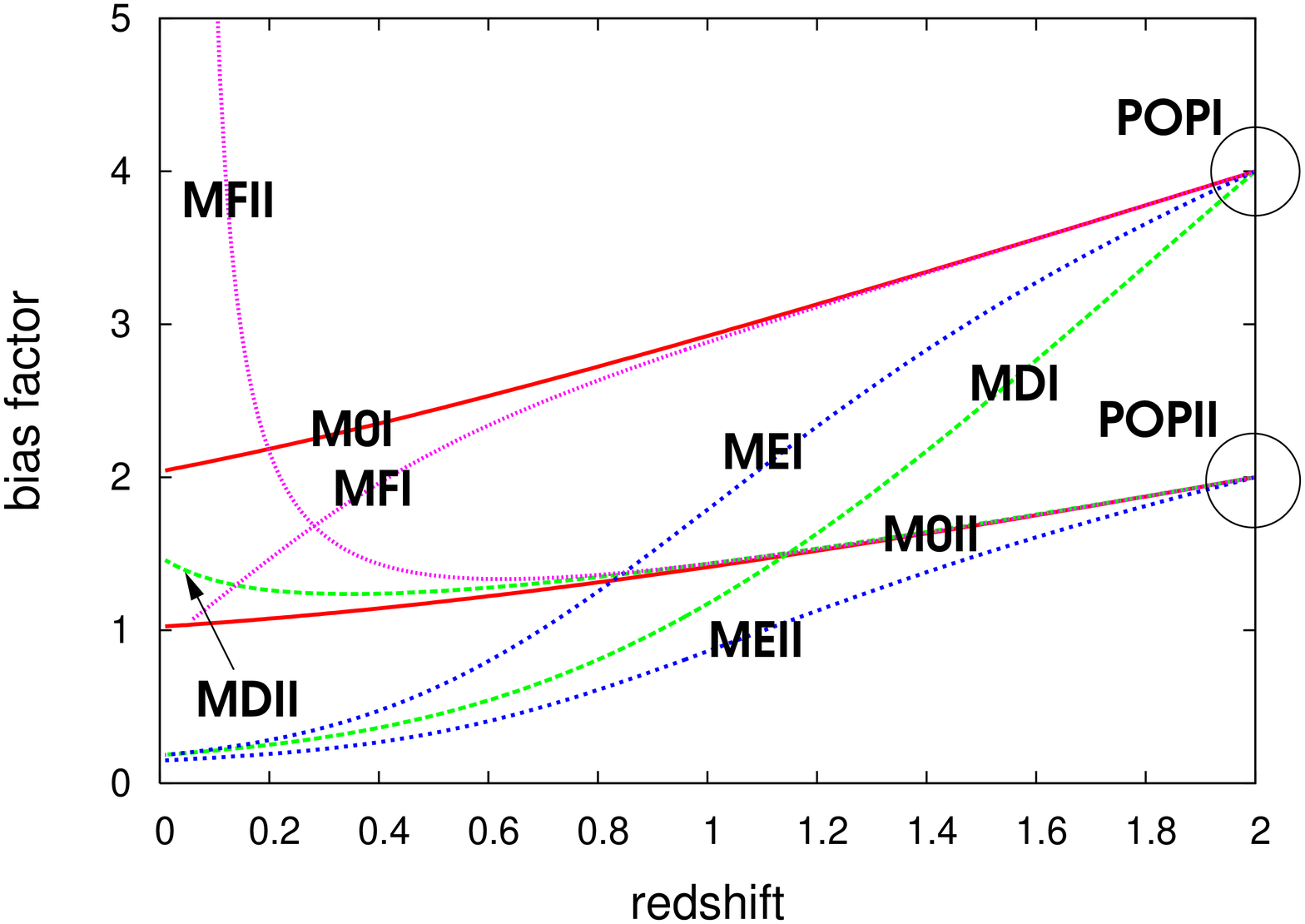,width=80mm,angle=0}
    \\
    \epsfig{file=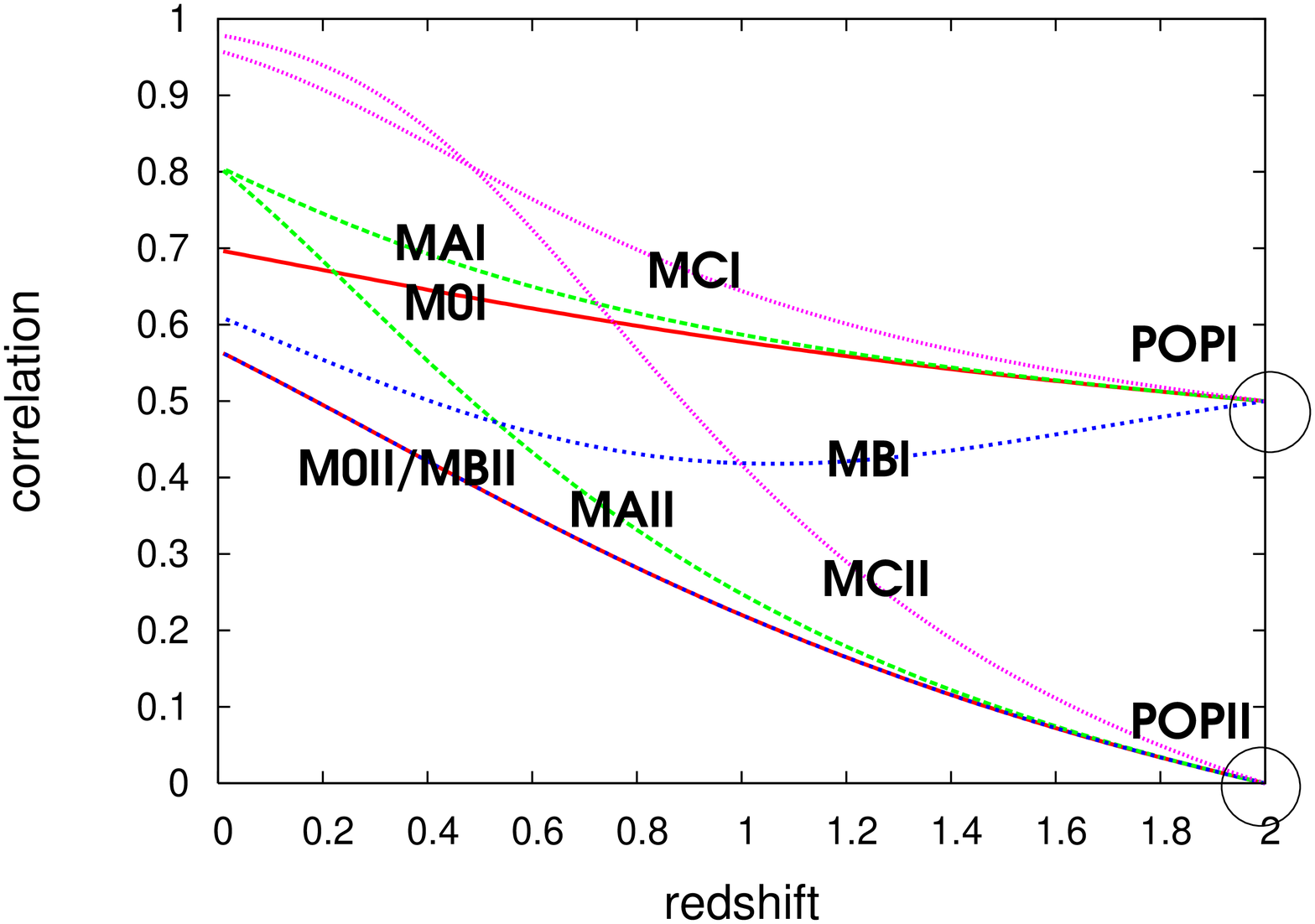,width=80mm,angle=0}&
    \epsfig{file=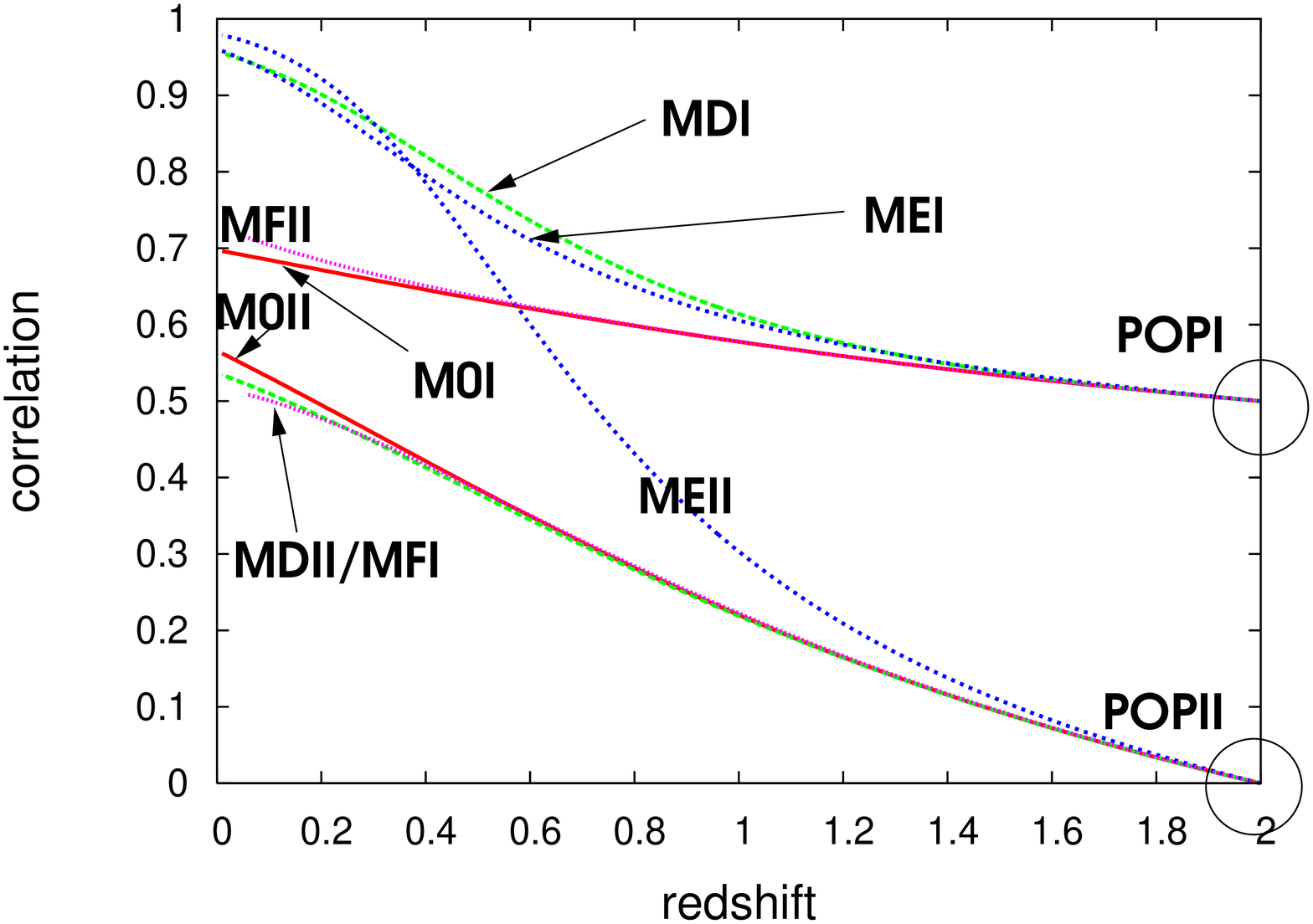,width=80mm,angle=0}
    \\
    \epsfig{file=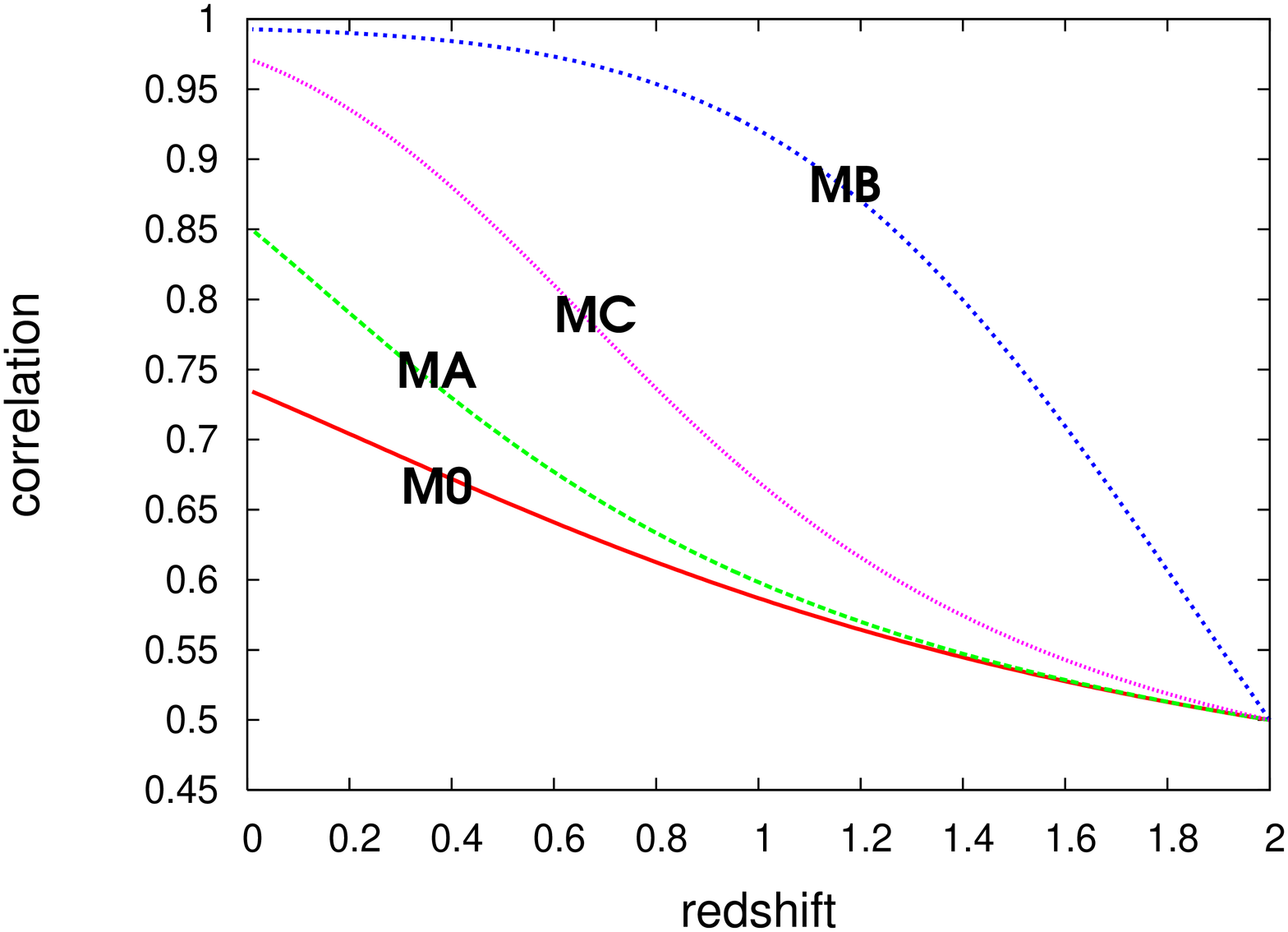,width=80mm,angle=0}&
    \epsfig{file=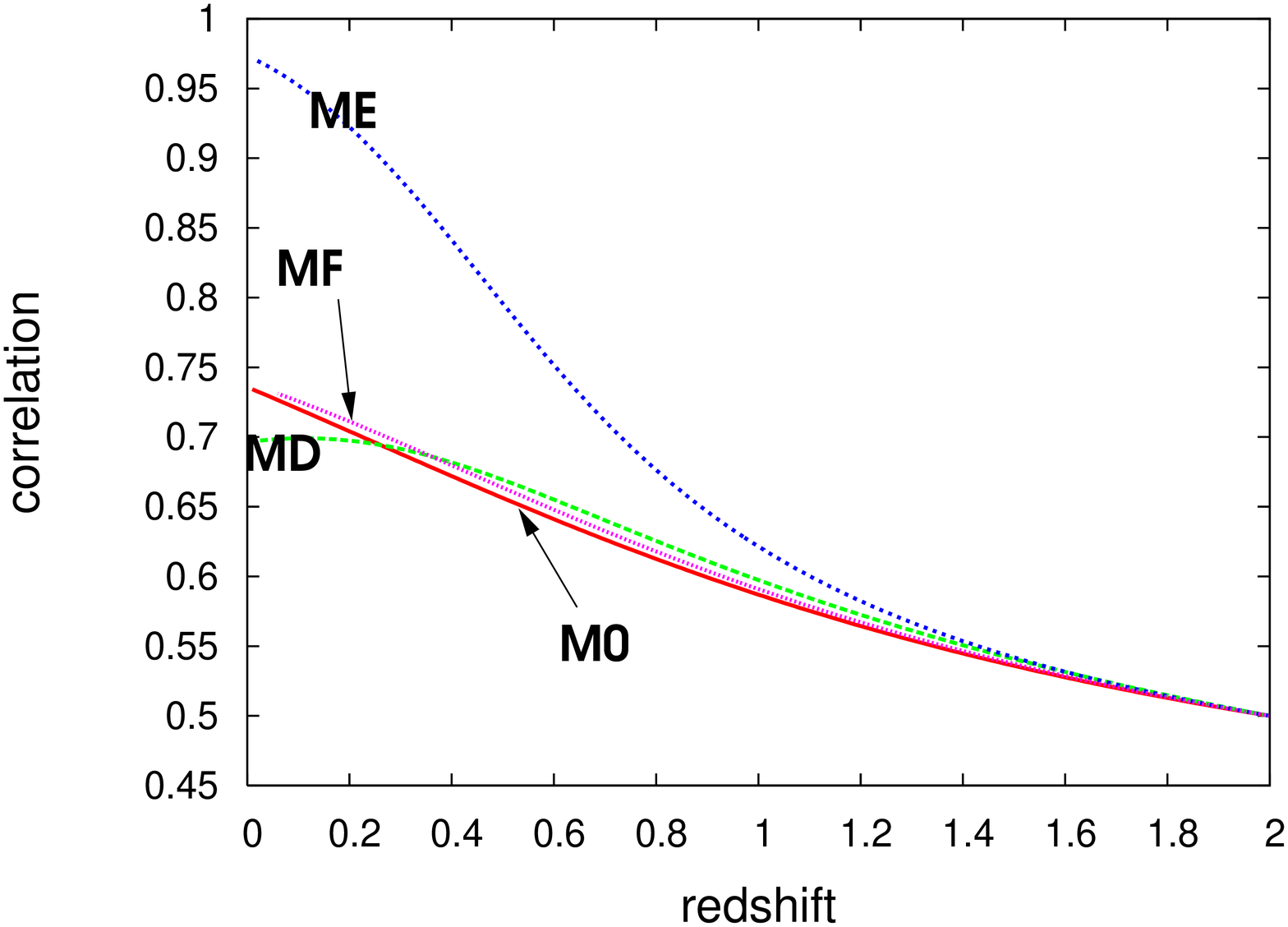,width=80mm,angle=0}
    \end{tabular}
  \end{center}
  \caption{
    Example evolutionary tracks of two galaxy populations POPI and
    POPII subject to different ``interactions''. All scenarios share
    same initial conditions at $z=2$ (POPI: $b_I=4$,$r_I=0.5$; POPII:
    $b_{II}=2$, $r_{II}=0$; $r_{I/II}=0.5$). Depicted are as a
    function of redshift (upper to lower row): bias factors $b_I$ and
    $b_{II}$ with respect to dark matter, correlations $r_I$ and
    $r_{II}$ to the dark matter field, and correlation $r_{I/II}$
    between the two galaxy fields (Sects. 4.2 and 4.3 for details)
    \newline \textbf{left column (linear couplings):} (arbitrary
    units) \textbf{M0}: interaction free case; \textbf{MA}: constant
    creation of galaxies with $A_I=A_{II}=10$; \textbf{MB}: POPII
    galaxies being transformed to POPI galaxies with
    $B_I^{II}=-B_{II}^{II}=10$; \textbf{MC}: linear coupling of both
    POPI and POPII to dark matter field with $C_I=C_{II}=10$ \newline
    \textbf{right column (quadratic couplings):} (arbitrary units)
    \textbf{M0}: interaction free evolution; \textbf{MD}:
    ``colliding'' POPII galaxies are transfered to POPI with
    $D_I^{II~II}=-D_{II}^{II~II}=10^{-4}$; \textbf{ME}: both
    populations are coupled to $\rho_m^2$ with $E_I=E_{II}=1$;
    \textbf{MF}: POPI galaxies are produced by $\Phi_I\propto{\rm
      n}_{II}\rho_m$ as much as POPII galaxies are destroyed,
    $F_I^{II}=-F_{II}^{II}=0.1$.}
  \label{linearfig}
\end{figure*}

\subsection{Linear couplings}

We first focus on the linear couplings by the $A_i$, $B_i^j$ and $C_i$
interaction terms. For these three scenarios (MA, MB and MC
respectively), we plot in Fig. \ref{linearfig} the evolutionary tracks
of the linear bias of two different galaxy populations.

The first population, hereafter POPI, has initially at redshift $z=2$
a bias factor $b_I=4$ and correlation $r_I=0.5$ with respect to the
dark matter. The second population, hereafter POPII, has $b_{II}=2$
and $r_{II}=0$ at $z=2$; it is thus initially not correlated to the
dark matter. The relative correlation between POPI and POPII we set to
$r_{I/II}=0.5$, well below maximum possible value of $r_{I/II}=0.87$
(according to Eq. \ref{inequ1}).  The number density of galaxies is
not constant due to the interaction (not plotted).

For the scenario MB, we assume that POPI is coupled to POPII such that
galaxies are transfered from POPII to POPI keeping the overall galaxy
number unchanged, thus $B_I^{II}=-B_{II}^{II}$.  Moreover, for that
particular scenario we increase the initial number of POPII galaxies
so that $\bar{n}_{II}=10\bar{n}_I$. In all other scenarios we used
$\bar{n}_I=\bar{n}_{II}$. Everywhere we use $\bar{\rm n}_I=1$ in
arbitrary units.

\subsection{Quadratic couplings}

For the toy models in this section, we assume that the bias parameters
are scale-independent, so that $\widehat{b_ib_jr_{ij}}=b_ib_jr_{ij}$
and $\widehat{b_jr_{ij}}=b_jr_{ij}$, where
$\left(b_i~b_{ij}~r_i~r_{ij}\right)$ are the large-scale bias
parameter as described in Sect. 2.4. Furthermore, we model the window
$\tilde{W}\left(k\right)$ (see Sect. 3) as a constant function with a cutoff
beyond a typical interaction scale $k_{\rm int}=2\pi/R_{\rm int}$,
here chosen to be $R_{\rm int}=1~\rm Mpc$.  We use the PD96
approximation for the non-linear evolution of the dark matter power
spectrum to estimate
\begin{equation}
  \ave{\delta_m^2}=
  \frac{1}{2\pi^2}\int_0^{k_{\rm int}} 
  dk~k^2~P_{\rm m}\left(k\right)
  \; .
\end{equation}
Fig. \ref{dmsqd} shows the estimates for different scales.

Fig. \ref{linearfig} shows examples of non-linear (quadratic)
couplings as conveyed by the interaction terms $D_i^{rs}$, $E_i$ and
$F_i^r$. These interactions lead to the scenarios MD, ME and MF
respectively. Again, as in the foregoing section, we have two galaxy
populations POPI and POPII with the aforementioned initial conditions.
For the mean galaxy density we set $\bar{n}_{II}=100\bar{\rm n}_I$,
except for ME where we assumed the same initial density for both
populations.

As before, we do not plot the evolution of the number densities. MD
couples POPI to POPII such that galaxies are added to POPI by
``collisions'' of POPII galaxies, while the same amount of galaxies is
taken from POPII ($D_I^{II~II}=-D_{II}^{II~II}$, all others are zero).
MF transfers galaxies from POPII to POPI by a quadratic coupling of
the dark matter and POPII density field, hence creating new POPI
galaxies everywhere where the density of both the dark matter and
POPII galaxies is high. Here, we also adjust the coupling constants
$F_i^j$ such that the overall galaxy density remains constant
($F_I^{II}=-F_{II}^{II}$).

\section{Discussion and conclusions}

Taking the hypothesis for granted that the bulk flow of galaxies is
identical to the bulk flow of the dark matter field, we derive a set
of differential equations that describe the evolution of the two-point
correlations between different galaxy populations and the dark matter
density field in terms of correlation power spectra (Eqs.
\ref{linpevolve1}, \ref{linpevolve2} and \ref{lindensity}).
Incorporated into this model is an ``interaction'' $\Phi_i$ that
allows for the destruction or creation of galaxies; in this paper, the
term interaction is used equivalently to a local change of the galaxy
number density. It may have explicit time dependence.

The model is valid only on scales where three point correlations of
all cosmological fields (density and velocity fields) are negligible.
This is fulfilled on large scales where the fields are Gaussian due to
the initial conditions of structure formation at high redshifts (as
seen in the CMB) and due to the fact that the field evolution is
essentially linear on large scales.  On small scales, this assumption
is definitely wrong, because gravitational instability has been
destroying Gaussianity proceeding gradually from smallest to larger
scales. The present stage of structure formation in the local Universe
is such that this transition form linear to non-linear scales occurs
at about $10$ Mpc $h^{-1}$; at earlier times, this scale was smaller.

We closely study an interaction rate $\Phi_i$ that is a local function
of the (smoothed) dark matter density field and the galaxy number density fields
up to second order; within the model, the choice of the interaction is
completely free though.  With this interaction, we introduce the
coupling constants $A_i$, $B_i^r$, $C_i$, $D_i^{rs}$, $E_i$ and
$F_i^r$ (see Eq. \ref{phidef}). Generally, this interaction term may
be pictured as the Taylor expansion of some complicated interaction
$\Phi_i\left({\rm n}_j,\rho_{\rm m}\right)$ up to second order. 
Nevertheless, some of the terms
associated with the coupling constants taken alone bear a simple 
interpretation. $D_i^{rs}$ may be used to describe interaction rates
of galaxy-galaxy collisions or mergers. Merging is an important
process in the currently favoured $\Lambda$CDM Universe
(e.g. Lacey~\&~Cole 1993). Linear couplings between the galaxy fields,
$B_i^r$, have a physical analogue as well: they describe
processes that transfer a certain fraction of one
galaxy population to another population  per volume and time, 
making the local
creation/destruction rate of galaxies proportional to the local
density of the other population (passive evolution). A 
constant production/destruction rate of galaxies, $A_i$, is just a
special case as it acts like a
linear coupling to a completely homogeneous field of galaxies.
\\
The 2nd order couplings between dark matter and galaxy fields, $E_i$
and $F_i^r$, and the linear coupling between dark matter and galaxies,
$C_i$, may be used, for instance, to describe formation processes that 
directly require the presence of dark matter overdensities, 
albeit the interpretation of these terms alone is less clear. At
least, one can say that linear couplings to the dark matter field
produce galaxies that are not biased with respect to the dark matter,
while a quadratic coupling makes relatively more galaxies in
overdensity regions.

General descriptions of a local stochastic bias like the one from
Dekel~\&~Lahav (1999) are based on the joint pdf of the (smoothed)
density contrasts of the considered fields.  Therefore - however the
defined bias parameters may look like - they have to be function of
the cumulants $\ave{\delta_{i_1}^{n_1}...\delta_{i_N}^{n_N}}_c$ of
this pdf, so that these are the basic quantities that should be
examined. Due to the Gaussianity of the fields on linear scales only
the second order cumulants are non-vanishing and hence only the
stochastic linear bias parameters in \Ref{linearbias} are relevant;
the first order cumulants vanish according to the definition of the
density contrasts.  Their evolution is described by means of Eqs.
\Ref{bi} to \Ref{linden}; Table \ref{tab} lists the interaction terms
based on the interaction correlators for the 2nd order Taylor
expansion of $\Phi_i$.  Our model distinguishes between the linear
bias $\left(b_i~r_i\right)$ of a galaxy population with respect to the
dark matter field and the linear bias $\left(b_{ij}~r_{ij}\right)$
between two galaxy populations.  The bias factor ``$b$'' can be
pictured as the ratio of the clustering strengths of the two fields,
whereas the correlation parameter ``$r$'' measures how strongly the
peaks and valleys of the density fields coincide. Note, however, that
also a possible non-linearity in the relation between $\delta_i$ and
$\delta_j$ affects the correlation parameter (Dekel~\&~Lahav
1999). On the large smoothing scales considered by this paper this is
neglectable though.

For all fields perfectly correlated to the dark matter field, thus
$r_i=r_{ij}=1$, the interaction terms $I^2_i$ and $I^3_{ij}$ always
vanish and therefore all correlations $\left(r_i~r_{ij}\right)$ are
``frozen in'' according to Eqs.  \Ref{ri} and \Ref{rij}. In that case,
the model reduces basically to Eq. \Ref{bi} and \Ref{linden} with all
correlations set to one ($b_{ij}=b_i/b_j$):
\begin{equation}
  \dt{b_i}=R\left(t\right)\frac{1-b_i}{2}+\left.I^1_i\right|_{r_i=r_{ij}=1}
  \; .
\end{equation}
The bias $b_i$ is then called \emph{deterministic}, since there is no
randomness in the relationship of the local density contrasts.  For
highly correlated fields, this can be a good approximation.

With no interaction present (Sect. 2.5), we obtain as TP98 and others
a debiasing of an initially biased galaxy field; this makes the galaxy
distribution looking more and more like the distribution of the
underlying dark matter distribution (Figs.  \ref{noninteractbias}).
The bias factor $b_i>1$ of a galaxy population that is less correlated
to the dark matter field declines faster than a more correlated
population (Fig.  \ref{noninteractrelbias}).  The figure also
demonstrates that differently correlated galaxy populations 
can temporarily evolve a relative bias factor 
$b_{ij}\ne 1$ with respect to each other
even though they may have had $b_{ij}=1$ at some time and they are not
interacting with each other. Moreover, characteristic
for an only slightly correlated population, $r_i<1$, is an
``overshoot'' that makes the population antibiased, i.e. $b_i<1$,
after some time. Later on, the bias factor increases again thereby
producing a relative minimum in $b_i$.  This minimum is clearly seen
in Fig.  \ref{noninteractrelbias}; according to Eq.  \Ref{bi} it has
to occur at the time where $r_i=b_i$, because $\dt{b_i}$ vanishes
there. On the other hand, this means that a possible local minimum of
$b_i$ always has to be smaller than one since $r_i\le 1$. In the
absence of any interaction, the relative correlation is a monotonic,
always increasing function; this is due to the rhs of Eq. \Ref{ri}
which always has a positive sign as long as $R\left(t\right)>0$.

\begin{figure}
  \begin{center}
    \epsfig{file=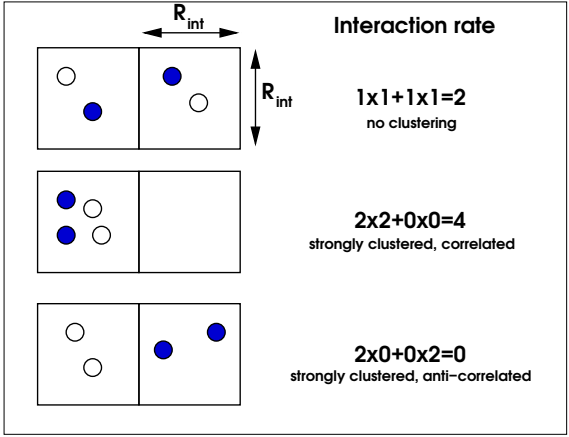,width=75mm,angle=0}
  \end{center}
  \caption{Sketch illustrating the effect of clustering and
    correlation on the mean interaction rate $\ave{\Phi}$, thus mean density
    evolution for quadratic couplings (``collisions'')
    $\Phi_{12}\propto n_1n_2$ between bright and dark particles. 
    The space is divided into two cells only whose
    size corresponds to some typical interaction distance $R_{\rm
    int}$; interactions take place between particles within the same cell.
    From top to bottom: homogeneous distribution, strong clustering
    and high correlation between particles, strong clustering and high
    anti-correlation between particles. Strong clustering and high
    correlation obviously results in the highest interaction rate and
    hence influences the mean density evolution the most.}
  \label{sketch}
\end{figure}

A few examples of linear couplings are plotted in Fig.
\ref{linearfig}. A linear coupling of a field ``II'' to a field ``I''
via $\Phi_I\propto n_{II}$ has the effect that the field of
\emph{newly formed or recently destroyed galaxies} type ``I'', ${\rm
  d}n_I=\Phi_I{\rm d}t$, has the same bias than the field of the
galaxies ``II''. In case of the formation of galaxies ``I'' (positive
sign in $\Phi_I$), this enriches the population ``I'' with new
galaxies having the same correlations as the galaxies in field ``II''.
Therefore, the bias factor between ``I'' and ``II'' is being reduced while
their correlation is being increased.  A positive linear coupling to the
dark matter field hence debiases a galaxy field quicker than without
interaction (like the populations POPI and POPII in scenario MC). The
linear coupling of POPII to POPI literally ``drags'' the population
POPI towards POPII as can be seen in MBI, while POPII (MBII), even
though loosing galaxies, shows the same behaviour as without
interaction in M0II; this is because it is linearly coupled to itself.
The interaction $A_i$ creates or destroys galaxies (depending on the
sign) with the same rate everywhere; this can be pictured as a linear
coupling to an absolutely homogeneous, fluctuation free field, having
$b=0$ with respect to any other field. $I^0_i$ for $B_i^j$ in table
\ref{tab} indeed reduces up to a constant to the $I^0_i$ for $A_i$, if
we set $b_j=0$. It is therefore not surprising that a constant
production of galaxies pulls the bias towards zero (see MAI and MAII),
more and more suppressing the density fluctuations.

In conclusion, a linear coupling of a field ``II'' to field ``I'' only
influences the bias evolution of ``I'' if ``II'' is biased with regard
to ``I''. In particular, a new population ``I'' being created solely
from a linear coupling to some other population ``II'' can never
become biased with respect to ``II''. Early type galaxies that may be
formed from spiral galaxies can therefore not be produced by a linear
coupling to the spiral galaxy field if they are biased with
  respect to spirals as observations imply (Norberg et al. 2002).
The fact that values for $\beta=\Omega_{\rm m}^{0.6}/b$ derived from
the IRAS (preferentially spiral galaxies) and the ORS (optically selected
galaxies) are consistent if a relative bias of $b_{\rm ORS}/b_{\rm
  IRAS}=1.4$ is assumed (Baker et al. 1998), also implies a bias
between spirals and ellipticals on large scales. If this is the case
then following the above arguments, ellipticals cannot simply be 
passively evolved spirals.

\begin{figure}
  \begin{center}
    \epsfig{file=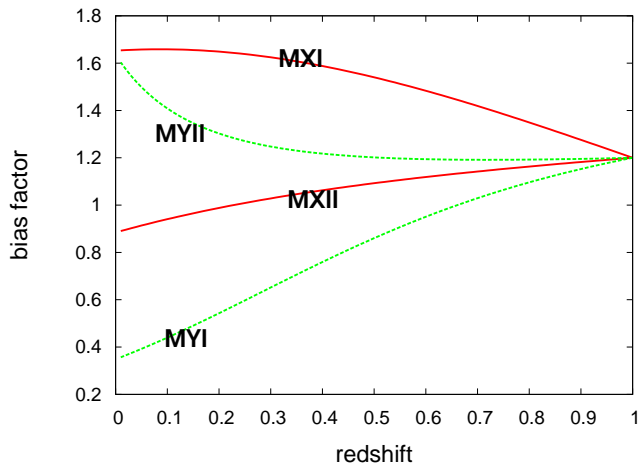,width=90mm,angle=0}
  \end{center}
  \caption{Evolutionary tracks of the bias factor with respect to the dark
    matter of two galaxy populations POPI and POPII for two scenarios
    MX and MY (initially at $z=1$: deterministic bias
    $b_I=b_{II}=1.2$, $\bar{n}_I=1$ and $\bar{n}_{II}=10^{-1}$).  Both
    scenarios assume that collisions of POPII galaxies create POPI
    with $D_I^{II~II}=-D_{II}^{II~II}\equiv D$; colliding galaxies are
    removed. \textbf{MX}: sets $\ave{\delta_{\rm m}^2}=0$ or
    equivalently $\widehat{b_I b_{II} r_{I/II}}=0$ 
    and $D=0.1$; \textbf{MY}: sets
    $\ave{\delta_{\rm m}^2}$ according to Fig. \ref{dmsqd} with $\rm
    R_{\rm int}=1$Mpc and $D=0.01$ in order to have roughly the same
    evolution of $\bar{n}_I$ and $\bar{n}_{II}$ as MX.}
  \label{dmsqddemo}
\end{figure}

Quadratic interactions, physically interpreted as collisions or
mergers, could do the job however. The reason is that the field of
newly formed or recently destroyed galaxies of type ``I'', coupled
quadratically to ``II'' is proportional to $n_{II}^2$.  These galaxies
have therefore the bias factor of $n_{II}^2$, which in general is
different from the bias factor of ``II''; in fact, the density
contrast $\delta_{\rm new}$ of the \emph{newly formed galaxies type ``I''} is then
\begin{equation}
  \delta_{\rm new}=
  \frac{n_{II}^2}{\ave{n_{II}^2}}-1=
  \frac{2\delta_{II}+\delta_{II}^2-\ave{\delta_{II}^2}}
  {1+\ave{\delta_{II}^2}}
\; .
\end{equation}
Smoothing $\delta_{\rm new}$ out to sufficiently large scales gives 
\begin{equation}
  \delta_{\rm new}\approx\frac{2\delta_{II}}{1+\ave{\delta_{II}^2}}
  \; ,
\end{equation}
because $\delta_{II}^2$ smoothed on large scales is approximately
$\ave{\delta_{II}^2}$ due to the ergodicity of the random field. Small
fluctuations $\ave{\delta_{II}^2}\ll 1$ 
make the newly formed galaxies biased with a bias factor of about
$b_{\rm new}\approx 2$ since $1+\ave{\delta_{\rm II}^2}\approx 1$ and
$\delta_{\rm new}\approx 2\delta_{\rm II}$. For non-negligible
fluctuations, on the other hand, this bias is roughly $b_{\rm new}=2/\left(1+\ave{\delta_{\rm
    II}^2}\right)$, thus taking a value between $0$ and $2$. This particular
example, as a side remark, demonstrates nicely that interactions on
very small scales can have an impact on the large-scale bias.
As an example, see Fig.
\ref{dmsqddemo}. Here we have a hypothetical population of galaxies
(POPI) unbiased with respect to another population (POPII). Collisions of POPII
galaxies add galaxies to POPI which then become biased or antibiased
depending on whether $\ave{\delta_{\rm II}^2}\ll 1$ (scenario MX) or 
$\ave{\delta_{\rm II}^2}\gg 1$ (scenario MY).

Quadratic interactions, Eq. $\Ref{lindensity}$, present a challenging problem
since one has to know the fluctuations of the density
fields on small scales, or equivalently (see Eq. \ref{smallscale}) the dispersion of the dark
matter $\ave{\delta_{\rm m}^2}$, the linear bias parameters
\begin{equation}
\hat{\Vector{b}}\equiv\left(\widehat{b_ib_jr_{ij}}~\widehat{b_ir_j}\right)
\end{equation}
and a window function $W\left(r\right)$; the linear bias
parameters $\hat{\Vector{b}}$ are the weighted means of the $k$-space bias
parameters over all scales $k$, or the (real space) bias parameter on a scale
defined by $W\left(r\right)$ (like in Eq. \ref{linearbias2}).
The window function actually \emph{defines} what is meant by fluctuations on
small scales by introducing a smoothing scale\footnote{%
Note, that such a scale is implicitly always assumed in order
to model the distribution of point like galaxies as a continuum as we
do in this paper.} $R_{\rm int}$ of the (real space) fields entering
$\Phi_i$, like for instance $n_i$ and $n_j$ in $\Phi_i\propto
n_in_j$. 
This scale is determined by the physical process underlying the
interaction and therefore lies presumably deep in the non-linear
regime. Why are these additional parameters needed for quadratic couplings?
This can be seen by the following argument.
One could think of the whole model volume being subdivided into
small cubic cells with side length $R_{\rm int}$; a cell contains $N_i=n_i
R_{\rm int}^3$ ``particles'' $n_i$.  Roughly speaking, the model predicts the
evolution of the correlations between the particle numbers $N_i$ of cells
which are far apart (large scale) and the mean number of particles $\ave{N_i}$
inside the cells taking into account the
gravitational field of the dark matter, its increasing clustering and
the background cosmology. The interaction term $\Phi_i$ changes the
number of galaxies $N_i$ inside a particular cell depending on the
number of galaxies and/or dark matter mass present in the same
cell by ${\rm d}N_i=R_{\rm int}^3\Phi_i{\rm d}t$.
For linear couplings $\Phi_i\propto n_j$,
the size of these cells does not have an impact on $\ave{\dot{N}_i}$ and
thus $\ave{\dot{n}_i}$ since $\ave{\Phi_i}$ depends only on the total number of particles
of all cells; hence $R_{\rm int}$ does not turn up in the 
model Eq. \Ref{linden}. For non-linear couplings, however,
the mean interaction rate indeed depends on how the particles are
distributed among the cells which is expressed by $\hat{\Vector{b}}$.

To be able to explore a toy model including quadratic couplings, we
made the assumption that the small-scale bias parameters are identical
to the bias parameters on large scales; we hence assumed no
scale-dependence for the linear bias. In fact, this is not what is
expected for some galaxy populations: on large scales early and late
type galaxies share approximately the same distribution (more galaxies
inside super-clusters, less in the voids outside), while on smaller,
cluster-scales early and late types are somehow anti-correlated as
seen in the density-morphology relation (Dressler et al. 1997).

The terms containing elements of $\hat{\Vector{b}}$ 
only have an impact if $\rm
R_{\rm int}$ is small enough making $\ave{\delta_{\rm m}^2}\geq 1$
(see Eq. \ref{deltamsqd})
\emph{and} if the correlations $\left|\widehat{b_ib_jr_{ij}}\right|$
and $\left|\widehat{b_ir_j}\right|$ are significantly different from
zero.  The strength of these terms can change the evolution of the
linear bias completely, as can be seen in Fig. \ref{dmsqddemo}.
There, galaxies of a population I (POPI) are created by the
collision/merger of galaxies of another population II (POPII).  The
difference between the models MX and MY is, that the former switches
off the $\ave{\delta_{\rm m}^2}$ terms while the other takes them into
account. Both scenarios predict the emergence of a bias of POPI
relative to POPII at $z=0$.  However, in MX we finally at $z=0$ have
$b_I>b_{II}$ while in MY we have $b_I<b_{II}$. This demonstrates that
for 2nd order interactions, the
evolution of the mean densities depends strongly on the homogeneity of
the ``soup'' of the interacting populations.  The mean density of a
completely homogeneous mixture of galaxies evolves slower than for a
mixture of galaxies with some substructure/clustering, if the
interacting populations are highly correlated (Fig. \ref{sketch}
for an illustration).  Therefore, to predict the bias evolution in the
context of quadratic interactions the knowledge of both
$\ave{\delta_{\rm m}^2}$ and the small-scale bias may be crucial.

Fitting the model presented in this paper to observed 
large-scale bias parameters with the intention to
look for quadratic couplings states therefore a practical problem: the
weighted bias $\hat{\Vector{b}}$ and $\ave{\delta_{\rm m}^2}$ are
required.  
The weighted bias parameters $\hat{\Vector{b}}$, however, are beyond
the scope of the model of this paper, since the model is valid only on
large scales. However, the knowledge of $\hat{\Vector{b}}$ is only
needed for the mean density evolution (see Eq.  \ref{linden}). In
practice, both the bias parameter and the galaxy number densities are,
at least principally, an observable. 
Therefore, this problem may be disarmed by directly
estimating $\bar{n}_i$ and $\ave{\Phi_i}=\dt{\bar{n}_i}$ through, for
instance, fitting generic functions to the observed mean galaxy number
density (polynomials, for example).
An estimate of the number density%
\footnote{We would like to remind the reader
here that the mean densities are comoving mean densities which for
number conserved populations is constant for all time.}%
, however,  requires the knowledge of the galaxy
luminosity function at different redshifts for every preferred galaxy
population which is a formidable task- but not impossible (Bell et al. 2004).
Measuring the scale-dependence of the bias parameters (e.g. Hoekstra
et al. 2002, H02) 
is here another option. The bias at and about the scale of maximum
weight $w\left(k\right)$ (see Eq. \ref{weighting}) could be used as
an estimate for $\hat{\Vector{b}}$ which then is inserted as a
constraint into the fitting procedure for the \emph{large-scale
  bias}; $\ave{\delta_{\rm m}^2}$ may be predicted using the PD96
prescription along with assumptions on $R_{\rm int}$.

Compared to TP98, we did not include a random component for the galaxy
formation (their Sect. 3); the production/destruction of galaxies is
always a deterministic function of the density fields. However, such
an random element could by included by a coupling to an
additional field that is only weakly or not all correlated to the dark
matter field. The effect of this is that a galaxy population gets more
and more polluted by newly formed galaxies that are not or only weakly
correlated to the dark matter. 
Thereby the relaxation to the dark matter field gets
retarded or even reverted.  This scenario has a physical analogy if
one imagines the newly formed galaxies as a condensate from a baryonic
matter field at places of high density but low temperature in order to
meet the Jeans criterion for self-collapse (White~\&~Frenk 1991;
White~\&~Rees 1978).  At early time, these places were inside dark
matter haloes; massive enough to attract the appropriate amount of
baryonic matter and to let it cool efficiently, thus at positions
highly correlated to the peaks of the dark matter density field. Later
on, however, the intergalactic medium probably got too hot inside the
haloes to form galaxies, so that the formation of galaxies may have
been shifted outside the highest density peaks.  Consequently, the
formation sites of later formed galaxies maybe have not been as much
correlated to the dark matter field as the sites of the galaxies made
earlier on (Blanton et al. 1999). The construction of Appendix D may
be used to mimic the behaviour of this baryonic field.

The practical application of this or similar models may be to work out
the relation between galaxy populations in terms of fundamental
coupling constants attached to the galaxies based on observations of
the bias evolution. This parameters may help to disentangle the zoo of
galaxy types and to reconstruct evolutionary paths.  Such observations
could be extracted, for instance, from weak gravitational lensing
surveys (H02) or from the redshift space
distortion in galaxy redshift surveys (Pen 1998).  In order to recover
the redshift evolution of the bias, it is however necessary to
subdivide the data set into redshift bins and even further into galaxy
population bins.  Considering that recent works (H02) focus on the
bias of the galaxies on the whole at one average redshift, it is clear
that this cannot be done with currently available data.


\begin{acknowledgements}
  I would like to thank Lindsay King and Peter Schneider for many
  helpful discussions and their comments concerning this paper.  This
  work was supported by the Deutsche Forschungsgemeinschaft under the
  Graduiertenkolleg 787.
\end{acknowledgements}


\appendix

\section{Correlations of convolved fields with a third field}

Here we calculate the ensemble average
$\avp{\left[\tilde{f}\ast\tilde{g}\right]\left(\Vector{k}\right)\tilde{h}^\ast\left(\Vector{k}^\prime\right)}$,
which is the correlator between the convolution of two random fields
with a third random field. We have
$\tilde{h}^\ast\left(\Vector{k}^\prime\right)=
\tilde{h}\left(\Vector{-k}^\prime\right)$, since we are exclusively
working with real number fields.  Writing out explicitly the
convolution of $\tilde{f}$ and $\tilde{g}$ gives
\begin{eqnarray}\nonumber
  &&
  \avp{\left[\tilde{f}\ast\tilde{g}\right]\left(\Vector{k}\right)
    \tilde{h}\left(\Vector{-k}^\prime\right)}
  \\\nonumber
  &=&
  \ave{
  \tilde{h}\left(\Vector{-k}^\prime\right)
  \int \frac{d\Vector{k}^\pprime}{\left(2\pi\right)^3}~
  \tilde{f}\left(\Vector{k}^\pprime\right)
  \tilde{g}\left(\Vector{k}-\Vector{k}^\pprime\right)}
  \\\nonumber
  &=&
  \int \frac{d\Vector{k}^\pprime}{\left(2\pi\right)^3}~
  \ave{
  \tilde{h}\left(\Vector{-k}^\prime\right)
  \tilde{f}\left(\Vector{k}^\pprime\right)
  \tilde{g}\left(\Vector{k}-\Vector{k}^\pprime\right)}
  \\\nonumber
  &=&
  \int\frac{d\Vector{k}^\prime}{\left(2\pi\right)^3}\int d\Vector{k}^\pprime~
   \delta_D\left(\Vector{k}-\Vector{k}^\prime\right)
  B_{123}\left(-\Vector{k}^\prime,\Vector{k}^\pprime,\Vector{k}-\Vector{k}^\pprime\right)
  \\\label{calculus}
  &=&
  \int \frac{d\Vector{k}^\prime}{\left(2\pi\right)^3}~
  B_{123}\left(-\Vector{k},\Vector{k}^\prime,\Vector{k}-\Vector{k}^\prime\right)
  \; ,
\end{eqnarray}
where $B_{123}$ is the bispectrum of $f$, $g$ and $h$. The only
assumption that has been made here is that the considered random
fields are homogeneous, for which holds
\begin{eqnarray}\nonumber
  &&
  \langle
  \tilde{h}\left(\Vector{k}\right)
  \tilde{f}\left(\Vector{k}^\prime\right)
  \tilde{g}\left(\Vector{k}^\pprime\right)
  \rangle
  \\
  &=&
  \left(2\pi\right)^3
  \delta_D\left(\Vector{k}+\Vector{k}^\prime+\Vector{k}^\pprime\right)
  B_{123}\left(\Vector{k},\Vector{k}^\prime,\Vector{k}^\pprime\right)
  \; .
\end{eqnarray}
For Gaussian fields the bispectrum vanishes, so that on linear scales
contributions from these correlators can be neglected.

\section{Model equations in terms of the linear bias parameter}

Here we are using the definitions \Ref{linbias} of the linear
stochastic bias parameter, the model Eqs. \Ref{linpevolve1},
\Ref{linpevolve2} and \Ref{lindensity} to explicitly write down
differential equations for the linear bias.

We start with the bias factor $b_i$ relative to the dark matter field:
\begin{eqnarray}\nonumber
  \dt{b_i}&=&
  \dt{}\sqrt{\frac{\pow{ii}}{\pow{m}}}=
  \frac{1}{2\sqrt{\pow{ii}\pow{m}}}\dt{\pow{ii}}-
  \frac{1}{2}\sqrt{\frac{\pow{ii}}{\pow{m}}}
  \frac{1}{\pow{m}}\dt{\pow{m}}
  \\
  &=&
  \frac{b_i}{2}
  \left(\frac{1}{\pow{ii}}\dt{\pow{ii}}-\frac{1}{\pow{m}}\dt{\pow{m}}\right)
  \; ,
 \end{eqnarray}
 where the definition of $R\left(t\right)$ in Eq. \Ref{defR} has been
 used. From Eq. \Ref{linpevolve2} we obtain
\begin{eqnarray}
  \frac{1}{\pow{ii}}\dt{\pow{ii}}&=&
  R\left(t\right)\frac{r_i}{b_i}+
  \frac{2}{\bar{\rho}_i}\left[
    \frac{\Re{\avp{\tilde{\Phi}_i\tilde{\delta}_i^\ast}}}{\pow{ii}}
    -\ave{\Phi_i}\right]
  \; ,
 \end{eqnarray}
 where $\Re{\left[x\right]}\equiv\frac{1}{2}\left(x+x^\ast\right)$
 denotes the real part of $x$.  Plugging in this expression into the
 previous equation we get
\begin{eqnarray}\nonumber
  \dt{b_i}&=&
  R\left(t\right)\frac{r_i-b_i}{2}+I^1_i
  \\\label{biappendix}
  I^1_i&\equiv&\frac{1}{\bar{\rho}_i}\left[
    \frac{\Re{\avp{\tilde{\Phi_i}\tilde{\delta}_i^\ast}}}{\pow{m}}\frac{1}{b_i}
    -\ave{\Phi_i}b_i\right]
  \; .
\end{eqnarray}
We proceed in a similar fashion for the correlation $r_i$ to the dark
matter field:
\begin{eqnarray}
  \dt{r_i}&=&
  \dt{}\frac{\pow{i}}{\sqrt{\pow{ii}\pow{m}}}
  \\\nonumber
  &=&
  \frac{1}{\sqrt{\pow{ii}\pow{m}}}\dt{\pow{i}}-
  \frac{\pow{i}}{2\sqrt{\pow{ii}\pow{m}}}\left[
    \frac{1}{\pow{ii}}\dt{\pow{ii}}+
    \frac{1}{\pow{m}}\dt{\pow{m}}\right]
  \\\nonumber
  &=&
  \frac{1}{b_i}\frac{1}{\pow{m}}\dt{\pow{i}}-
  \frac{r_i}{2}\frac{1}{\pow{ii}}\dt{\pow{ii}}-
  R\left(t\right)\frac{r_i}{2}
  \; .
\end{eqnarray}
Now we need Eq. \Ref{linpevolve1} to go further:
\begin{eqnarray}
  \frac{1}{\pow{m}}\dt{\pow{i}}&=&
  R\left(t\right)\frac{1+r_ib_i}{2}+
  \frac{1}{\bar{\rho}_i}\frac{\avp{\tilde{\Phi}_i\tilde{\delta}_m^\ast}}{\pow{m}}-
  \frac{\pow{i}}{\pow{m}}\frac{\ave{\Phi_i}}{\bar{\rho}_i}
  \\\nonumber
  &=&
  R\left(t\right)\frac{1+r_ib_i}{2}+
  \frac{1}{\bar{\rho}_i}\left[
    \frac{\avp{\tilde{\Phi}_i\tilde{\delta}_m^\ast}}{\pow{m}}-
    b_ir_i\ave{\Phi_i}\right]
  \; .
\end{eqnarray}
Plugging these in yields for the correlation parameter
\begin{eqnarray}\nonumber
 \dt{r_i}&=&
  \frac{R\left(t\right)}{2}\frac{1-r_i^2}{b_i}+I^2_i
  \\\label{riappendix}
  I^2_i&\equiv&\frac{1}{\bar{\rho}_i}\left[
    \frac{\avp{\tilde{\Phi}_i\tilde{\delta}_m^\ast}}{\pow{m}}\frac{1}{b_i}-
    \frac{\Re{\avp{\tilde{\Phi}_i\tilde{\delta}_i^\ast}}}{\pow{m}}\frac{r_i}{b_i^2}
  \right]
\; .
\end{eqnarray}
Now we turn to the evolution of the linear bias parameter between two
galaxy populations, starting off with the bias $b_{ij}$:
\begin{equation}
  \dt{b_{ij}}=\dt{}\frac{b_i}{b_j}=
  b_{ij}\left[\frac{1}{b_i}\dt{b_i}-\frac{1}{b_j}\dt{b_j}\right]
  \; .
\end{equation}
The expressions in the bracket are worked out using Eq. \Ref{biappendix} so
that we therefore obtain
\begin{equation}
  \dt{b_{ij}}=
  \frac{R\left(t\right)}{2}~
  \frac{r_ib_j-r_jb_i}{b_ib_j}b_{ij}+
  \frac{b_{ij}}{b_i}I^1_i-\frac{b_{ij}}{b_j}I^1_j
   \; .
\end{equation}
The correlation $r_{ij}$ between two galaxy populations is derived in
the same way but in the end a bit lengthy:
\begin{eqnarray}
  \dt{r_{ij}}&=&
  \dt{}\frac{\pow{ij}}{\sqrt{\pow{ii}\pow{jj}}}
  \\\nonumber
  &=&
  r_{ij}\frac{1}{\pow{ij}}\dt{\pow{ij}}-
  \frac{1}{2}r_{ij}
  \left[
    \frac{1}{\pow{ii}}\dt{\pow{ii}}+
    \frac{1}{\pow{jj}}\dt{\pow{jj}}
  \right]
  \; .
\end{eqnarray}
The expressions in the bracket have been worked out before, so that
the only remaining unknown expression is (uses Eq. \ref{linpevolve2})
\begin{eqnarray}
  \frac{1}{\pow{ij}}\dt{\pow{ij}}&=&
  \frac{R\left(t\right)}{2}~\frac{b_ir_i+b_jr_j}{b_ib_jr_{ij}}
  \\\nonumber
  &+&
  \frac{1}{\bar{\rho}_i}
  \left[\frac{\avp{\tilde{\Phi}_i\tilde{\delta}_j^\ast}}{\pow{ij}}-\ave{\Phi_i}\right]
  +
  \frac{1}{\bar{\rho}_j}
  \left[\frac{\avp{\tilde{\Phi}_j^\ast\tilde{\delta}_i}}{\pow{ij}}-\ave{\Phi_j}\right]
  \; .
\end{eqnarray}  
Taking this into account, we finally get
\begin{eqnarray}\nonumber
  \dt{r_{ij}}&=&
  \frac{R\left(t\right)}{2}~
  \frac{\left(r_i-r_{ij}r_j\right)b_i+\left(r_j-r_{ij}r_i\right)b_j}{b_ib_j}
  +I^3_{ij}+\left[I^3_{ji}\right]^\ast
  \\
  I^3_{ij}&\equiv&\frac{1}{\bar{\rho}_i}\left[
    \frac{\avp{\tilde{\Phi}_i\tilde{\delta}_j^\ast}}{\pow{m}}\frac{1}{b_ib_j}-
    \frac{\Re{\avp{\tilde{\Phi}_i\tilde{\delta}_i^\ast}}}{\pow{m}}\frac{r_{ij}}{b_i^2}
  \right]
  \; .
\end{eqnarray}

The interaction rates $\Phi_i$ and the density contrasts $\delta_{\rm
  X}$ are real numbers, so that the correlators
$\avp{\tilde{\Phi}_i\tilde{\delta}_{\rm X}^\ast}$ have to be real
numbers too. For that reason, we are allowed to omit the real part
operator ``$\Re{}$'' in the interaction terms $I_i^1$, $I_i^2$ and
$I_{ij}^3$ as has been done in Eqs. \Ref{interactcorr}.

\section{Interaction correlators for first and second order $\Phi_i$}

As we are working with the density contrasts $\delta_i$ instead of the
densities ${n}_i$ itself, we rewrite the above expression for $\Phi_i$
in Eq. \Ref{phidef} using the definition \Ref{densitycontrast}:
\begin{eqnarray}
 \Phi_i&=&A_i+B_i^r\bar{n}_r+\hat{C}_i\bar{\rho}_{\rm m}
 \\\nonumber
 &+&D_i^{rs}\bar{n}_r\bar{n}_s+\hat{E}_i\bar{\rho}_{\rm
 m}^2+\hat{F}_i^r\bar{n}_r\bar{\rho}_{\rm m}
 \\\nonumber
 &+&B_i^r\bar{n}_r\delta_r+\hat{C}_i\bar{\rho}_{\rm m}\delta_{\rm
 m}+D_i^{rs}\bar{n}_r\bar{n}_s\left(\delta_r+\delta_s\right)
\\\nonumber
&+&2\hat{E}_i\bar{\rho}_{\rm m}^2\delta_{\rm
 m}+\hat{F}_i^r\bar{\rho}_{\rm m}\bar{n}_r\left(\delta_{\rm m}+\delta_r\right)
\\\nonumber
&+&D_i^{rs}\bar{n}_r\bar{n}_s
\delta_r\delta_s+\hat{E}_i\bar{\rho}_{\rm m}^2\delta_{\rm
 m}^2+\hat{F}_i^r\bar{n}_r\bar{\rho}_{\rm m}\delta_{\rm m}\delta_r
 \; .
\end{eqnarray}
Where possible, we absorb for convenience all $\bar{\rho}_{\rm m}$
inside the associated coupling constant, removing the previously
introduced hat ``~$\hat{}$~''. This absorption makes sense, because
$\bar{\rho}_m$ is supposed to be a constant and therefore produces in
this formalism a degeneracy between $\bar{\rho}_m$ and its associated
coupling constant. This results in
\begin{eqnarray}\label{sinksource}
 \Phi_i&=&A_i+C_i+E_i+\left(B_i^r+F_i^r\right)\bar{n}_r+
 D_i^{rs}\bar{n}_r\bar{n}_s
 \\\nonumber
 &+&B_i^r\bar{n}_r\delta_r+C_i\delta_{\rm m}+
 D_i^{rs}\bar{n}_r\bar{n}_s\left(\delta_r+\delta_s\right)+2E_i\delta_{\rm m}
 \\\nonumber
 &+&F_i^r\bar{n}_r\left(\delta_{\rm
 m}+\delta_r\right)+D_i^{rs}\bar{n}_r\bar{n}_s\delta_r\delta_s+
E_i\delta_{\rm m}^2+F_i^r\bar{n}_r\delta_{\rm m}\delta_r
 \; .
\end{eqnarray}
The Fourier transform of the interaction term is thus, throwing away
the terms contributing only at $\Vector{k}=0$
\begin{eqnarray}\nonumber
\tilde{\Phi}_i&=&
B_i^r\bar{n}_r\tilde{\delta}_r+C_i\tilde{\delta}_{\rm m}+
D_i^{rs}\bar{n}_r\bar{n}_s\left(\tilde{\delta}_r+\tilde{\delta}_s\right)+
2E_i\tilde{\delta}_{\rm m}
\\\nonumber
&+&
F_i^r\bar{n}_r\left(\tilde{\delta}_{\rm m}+\tilde{\delta}_r\right)+
D_i^{rs}\bar{n}_r\bar{n}_s\left(\tilde{\delta}_r\ast\tilde{\delta}_s\right)
\\\label{basicphi}
&+&
E_i\left(\tilde{\delta}_{\rm m}\ast\tilde{\delta}_{\rm m}\right)
+F_i^r\bar{n}_r\left(\tilde{\delta}_{\rm m}\ast\tilde{\delta}_r\right)
\; .
\end{eqnarray}

The model equations \Ref{didm2} and \Ref{didj2} require the
interaction correlators $\avp{\tilde{\Phi}_i\tilde{\delta}_m^\ast}$,
$\avp{\tilde{\Phi}_j\tilde{\delta}_i^\ast}$ and
$\avp{\tilde{\Phi}_i\tilde{\delta}_j^\ast}$ to be evaluated. The last
two are, of course, the same up to an exchange of the indices, so that
we only have to determine the first two. Using the definition of the
correlation power spectra in \Ref{powerdef} and the restriction to
Gaussian fields (bispectra emerging according to Appendix A are zero),
we obtain:
\begin{eqnarray}\label{intcorr}
  \avp{\tilde{\Phi}_i\tilde{\delta}_{\rm m}^\ast}
  &=&
  B_i^r\bar{n}_rP_r+C_iP_{\rm m}+
  D_i^{rs}\bar{n}_r\bar{n}_s\left(P_r+P_s\right)
  \\\nonumber
  &+&
  2E_iP_{\rm m}+F_i^r\bar{n}_r\left(P_{\rm m}+P_r\right)
  \\\nonumber\\\nonumber
  \avp{\tilde{\Phi}_i\tilde{\delta}_j^\ast}
  &=&
  B_i^r\bar{n}_rP_{rj}+C_iP_j+
  D_i^{rs}\bar{n}_r\bar{n}_s\left(P_{rj}+P_{sj}\right)
  \\\nonumber
  &+&
  2E_iP_j+F_i^r\bar{n}_r\left(P_j+P_{rj}\right)
  \\\nonumber\\\nonumber
  \avp{\tilde{\Phi}_j^\ast\tilde{\delta}_i}
  &=&
  B_j^r\bar{n}_rP_{ri}+C_jP_i+
  D_j^{rs}\bar{n}_r\bar{n}_s\left(P_{ri}+P_{si}\right)
  \\\nonumber
  &+&
  2E_jP_i+F_j^r\bar{n}_r\left(P_i+P_{ri}\right)
  \; .
\end{eqnarray}

Eq. \Ref{linden} for the mean density evolution, however, needs
the ensemble average of the interaction term in \emph{real space}
$\ave{\Phi_i}$. Doing so and removing terms linear in the density
contrasts due to $\ave{\delta_i}=0$, this results in
\begin{eqnarray}
 \ave{\Phi_i}&=&
 A_i+C_i+E_i+\left(B_i^r+F_i^r\right)\bar{n}_r+D_i^{rs}\bar{n}_r\bar{n}_s
 \\\nonumber
 &+&
 D_i^{rs}\bar{n}_r\bar{n}_s\ave{\delta_r\delta_s}+E_i\ave{\delta_{\rm m}^2}
 +F_i^r\bar{n}_r\ave{\delta_{\rm m}\delta_r}
 \; .
\end{eqnarray}

\section{Fields with constant bias}

Here we consider a new class of density fields - \emph{static fields}
- that may serve as a model source for producing galaxies.  Their
difference to the already described fields $\delta_i$ in Sect. 2 is
that they are supposed to have a constant bias with respect to the
dark matter \emph{for all time}; they are therefore some sort of random
component $\delta_\perp$ as in TP98. They are introduced therein in order
to serve as a source for creating new galaxies with a certain fixed
bias at the time of there formation. In contrast to the random
component in TP98, the static fields here do not necessarily
have to be totally uncorrelated to the dark matter field and do not
have to be coupled linearly only; hence the static fields are a bit
more general.

As we force this class of fields to have a constant bias relative to
the dark matter, they certainly do not obey Eq.  \Ref{basiceqfourier}
and hence have to be treated differently compared to the common galaxy
fields. As before, we restrict ourselves to the linear regime.  To
avoid confusion with the already studied fields, we use Greek letters
as indices, like for example $\delta_\alpha$ and
$\tilde{\delta}_\alpha$ for its Fourier coefficients.

Demanding the linear bias parameter $b_\alpha$ and $r_\alpha$ to be
constant, immediately fixes the equations for the correlation power
spectra $\pow{\alpha\alpha}$ and $\pow{\alpha}$ by virtue of the
definition \Ref{linbias}:
\begin{eqnarray}
  \dt{b_\alpha}=0&\Rightarrow&
  \dt{\pow{\alpha\alpha}}=b_\alpha^2\dt{\pow{m}}
  \\\nonumber
  \dt{r_\alpha}=0&\Rightarrow&
  \dt{\pow{\alpha}}=b_\alpha r_\alpha\dt{\pow{m}}
  \; .
\end{eqnarray}

The cross-correlation of $\delta_\alpha$ with one of the conventional
galaxy number density fields $\delta_i$ (Sect. 2) is not equally
obvious to the eye.  Since the bias relative to the dark matter stays
constant, we know that fluctuations of the static fields have to grow
with the same rate as the dark matter fluctuations
\begin{equation}
  \dt{\delta_\alpha}=\frac{R\left(t\right)}{2}\delta_\alpha
  \; ,
\end{equation}
where $R\left(t\right)$ is the rate of structure growth on linear
scales (Eq. \ref{defR}).  This relation yields
\begin{eqnarray}
  \dt{\pow{i\alpha}}&=&
  \dt{}\avp{\tilde{\delta}_i\tilde{\delta}_\alpha^\ast}
  =\avp{\dt{\tilde{\delta}_i}\tilde{\delta}_\alpha^\ast}+
  \avp{\tilde{\delta}_i\dt{\tilde{\delta}_\alpha^\ast}}
  \\\nonumber
  &=&
  \frac{R\left(t\right)}{2}\left(\pow{\alpha}+\pow{i\alpha}\right)
  +\frac{1}{\bar{n}_i}\left[\avp{\tilde{\Phi}_i\delta_\alpha^\ast}-
    \pow{i\alpha}\ave{\Phi_i}\right]
  \; ,
\end{eqnarray}  
where Eq. \Ref{basiceqfourier} for $\dt{\tilde{\delta}_i}$ has been
used (as usual bispectra terms have been neglected: Appendix A).

Analogue to Appendix B we then have
\begin{eqnarray}
  \dt{b_{i\alpha}}&=&
  \dt{}\frac{b_i}{b_\alpha}=\frac{1}{b_\alpha}\dt{b_i}
  \\\nonumber
  &=&
  \frac{R\left(t\right)}{2}~\frac{r_i-b_i}{b_\alpha}
  +\frac{1}{b_\alpha}I^1_i
  \; ,
\end{eqnarray}
and
\begin{eqnarray}\nonumber
  \dt{r_{i\alpha}}&=&
  r_{i\alpha}\frac{1}{\pow{i\alpha}}\dt{\pow{i\alpha}}-
  \frac{1}{2}r_{i\alpha}
  \left[
    \frac{1}{\pow{ii}}\dt{\pow{ii}}+
    \frac{1}{\pow{\alpha\alpha}}\dt{\pow{\alpha\alpha}}
  \right]  
  \\
  &=&
  \frac{R\left(t\right)}{2}~\frac{r_\alpha-r_ir_{i\alpha}}{b_i}+
  I^3_{i\alpha}
  \; ,
\end{eqnarray}
where the definitions of
$I^3_{i\alpha}=\left.I^3_{ij}\right|_{j=\alpha}$ and $I^1_i$ in
Appendix B have been used.

Setting $b_\alpha=r_\alpha=1$ and $r_{i\alpha}=r_i$ reduces
$\dt{r_{i\alpha}}$ and $\dt{b_{i\alpha}}$ to $\dt{b_i}$ (Eq. \ref{bi})
and $\dt{r_i}$ (Eq. \ref{ri}) respectively. This tells us that the
dark matter field is just a special case of the here introduced static
fields, since it (trivially) stays unbiased with respect to itself all
the time.


\end{document}